\documentclass[longauth]{aa}  
\usepackage{natbib}
\usepackage{graphicx}
\usepackage{txfonts}
\usepackage{hyperref}
\usepackage{lscape}
\usepackage{xcolor}
\usepackage{ulem}
\usepackage{color, soul} %Pedro: I put this here
\def\ms{\hbox{\,m\,s$^{-1}$}}         %m.s -1
\def\cms{\hbox{\,cm\,s$^{-1}$}}       %cm.s -1
\def\kms{\hbox{\,km\,s$^{-1}$}}       %km.s -1
\def\sini{\hbox{sin\,$i$}}      %vsini
\def\Msun{\hbox{$\mathrm{M}_{\odot}$}}             %Msun
\def\Rsun{\hbox{$\mathrm{R}_{\odot}$}}
\def\Mjup{\hbox{$\mathrm{M}_{\rm Jup}$}}
\def\Rjup{\hbox{$\mathrm{R}_{\rm Jup}$}}

\def\mearth{\hbox{M$_\oplus$}}
\def\rearth{\hbox{R$_\oplus$}}
\begin{document} 
  \title{A precise architecture characterization of the \object{$\pi$\,Men} planetary system\thanks{Based (in part) on Guaranteed Time Observations collected at the European Southern Observatory under ESO programme(s) 1102.C-0744, 1102.C-0958 and 1104.C-0350 by the ESPRESSO Consortium. Tables B.1 and B.2 are only available in electronic form at the CDS via anonymous ftp to cdsarc.u-strasbg.fr (130.79.128.5) or via http://cdsweb.u-strasbg.fr/cgi-bin/qcat?J/A+A/}
  } 

  \author{M.\,Damasso \inst{1}
  \and A.\,Sozzetti \inst{1}
  \and C.\,Lovis \inst{2}
  \and S.\,C.\,C.\,Barros \inst{3}
  \and S.\,G.\,Sousa \inst{3}
  \and O.\,D.\,S.\,Demangeon \inst{3}
  \and J.\,P.\,Faria \inst{3,4}
  \and J.\,Lillo-Box \inst{5}
  \and S.\,Cristiani\inst{6,7}
  \and F. Pepe\inst{2}
% ...
}

\institute{INAF -- Osservatorio Astrofisico di Torino, Via Osservatorio 20, I-10025 Pino Torinese, Italy
\email{mario.damasso@inaf.it}
\and D\'epartement d'astronomie, Universit\'e de Genève,51, ch. des Maillettes, CH-1290 Versoix
\and Instituto de Astrof\'isica e Ci\^encias do Espa\c{c}o, Universidade do Porto, CAUP, Rua das Estrelas, 4150-762 Porto, Portugal
\and Departamento de F\'isica e Astronomia, Faculdade de Ci\^encias, Universidade do Porto, Rua do Campo Alegre, 4169-007 Porto, Portugal
\and Centro de Astrobiolog\'ia (CSIC-INTA), Carretera de Ajalvir km 4, E-28850 Torrej\'on de Ardoz, Madrid, Spain
\and INAF -- Osservatorio Astronomico di Trieste, Via Tiepolo 11, I-34143 Trieste, Italy
\and Institute for Fundamental Physics of the Universe, IFPU, Via Beirut 2, 34151 Grignano, Trieste, Italy
% ...
}
    
  \date{Received May 14, 2020; accepted }

  \abstract
  % context heading (optional)
  % {} leave it empty if necessary  
   {The bright star \object{$\pi$\,Men} was chosen as the first target for a radial velocity follow-up to test the performance of ESPRESSO, the new high-resolution spectrograph at the ESO's Very-Large Telescope (VLT). The star hosts a multi-planet system (a transiting 4 $\mearth$ planet at $\sim$0.07 au, and a sub-stellar companion on a $\sim$2\,100-day eccentric orbit) which is particularly appealing for a precise multi-technique characterization.}
  % aims heading (mandatory)
   {With the new ESPRESSO observations, that cover a time span of 200 days, we aim to improve the precision and accuracy of the planet parameters and search for additional low-mass companions. We also take advantage of new photometric transits of \object{$\pi$\,Men\,c} observed by \textit{TESS} over a time span that overlaps with that of the ESPRESSO follow-up campaign.}
  % methods heading (mandatory)
   {We analyse the enlarged spectroscopic and photometric datasets and compare the results to those in the literature. We further characterize the system by means of absolute astrometry with Hipparcos and Gaia. We used the high-resolution spectra of ESPRESSO for an independent determination of the stellar fundamental parameters.}
  % results heading (mandatory)
   {We present a precise characterization of the planetary system around $\pi$\,Men. The ESPRESSO radial velocities alone (37 nightly binned data with typical uncertainty of 10 cm/s) allow for a precise retrieval of the Doppler signal induced by $\pi$ Men\,c. The residuals show an RMS of 1.2 \ms, which is half that of the HARPS data and, based on them, we put limits on the presence of additional low-mass planets (e.g. we can exclude companions with a minimum mass less than $\sim$2 $\mearth$ within the orbit of $\pi$ Men\,c). We improve the ephemeris of $\pi$ Men\,c using 18 additional \textit{TESS} transits, and in combination with the astrometric measurements, we determine the inclination of the orbital plane of $\pi$ Men\,b with high precision ($i_{b}=45.8^{+1.4}_{-1.1}$\,deg). This leads to the precise measurement of its absolute mass $m_{b}=14.1^{+0.5}_{-0.4}$ \Mjup, indicating that $\pi$ Men\,b can be classified as a brown dwarf.} %This is in agreement with the non detection of the transit of planet b, that was expected to occur within the \textit{TESS} Observation Sector 12 based on our prediction.}
  % conclusions heading (optional), leave it empty if necessary 
   {$\pi$\,Men represents a nice example of the extreme precision radial velocities that can be obtained with ESPRESSO for bright targets. Our determination of the 3-D architecture of the \object{$\pi$\,Men} planetary system, and the high relative misalignment of the planetary orbital planes, put constraints and challenges to the theories of formation and dynamical evolution of planetary systems. The accurate measurement of the mass of $\pi$ Men\,b contributes to make the brown dwarf desert a bit greener. }

   \keywords{Techniques: radial velocities; photometric -- Astrometry -- Planetary systems -- Stars: individual: $\pi$\,Men; HD 39091; TOI-144}

   \maketitle
%
%________________________________________________________________
\section{Introduction}

The Southern Hemisphere bright star \object{$\pi$\,Men} (HD 39091; V=5.7 mag, spectral type G0V) became a high-priority target for follow-up with high-precision spectrographs after the NASA Transiting Exoplanet Survey Satellite (\textit{TESS}; \citealt{ricker15}) detected the transiting super-Earth/sub-Neptune \object{$\pi$\,Men\,c} ($P_{\rm c}$ $\sim$6.27 days; $R_{\rm c}$ $\sim$2 $\rearth$). This was one of the most relevant among the first discoveries of \textit{TESS} after it started scientific observations at the end of July 2018 (\object{$\pi$\,Men} is also known as \textit{TESS} object of interest TOI-144). Following the discovery announcement, \cite{huang18} and \citet{gandolfi18} independently detected the spectroscopic orbit of \object{$\pi$\,Men\,c} by analysing archival radial velocities (RVs) of HARPS and UCLES, and confirmed its planetary nature. 
The brightness of the star made it a perfect target for testing the performance of the new-generation ultra-stable and high-resolution \'Echelle SPectrograph for Rocky Exoplanets and Stable Spectroscopic Observations (ESPRESSO; \citealt{Pepe-2020}) of ESO's Very-Large Telescope (VLT). In fact, \object{$\pi$\,Men} was one of the first science target observed with ESPRESSO, with the aim to use the very precise radial velocities ($\sigma_{\rm RV}\sim$10 $\cms$) to improve the measurement of the mass and bulk density of \object{$\pi$\,Men\,c}. 

The determination of precise planetary physical properties is crucial for successive investigations of a planet's atmosphere, in particular for a strongly irradiated super-Earth-to-sub-Neptune-sized planet like \object{$\pi$\,Men\,c}. Following \cite{batalha19}, a precision in the mass determination of at least $20\%$ is required for detailed atmospheric analyses through transmission spectroscopy, as those that will be made possible with the James Webb Space Telescope. The growing interest around \object{$\pi$\,Men\,c} has led \cite{king2019} to present it as a favourable target to search for ultraviolet absorption due to an escaping atmosphere and, right after, \cite{munoz2020} announced the non detection of neutral hydrogen in the atmosphere of the planet through Lyman-$\alpha$ transmission spectroscopy with the Hubble Space Telescope. They show that the lack of an extended atmosphere would make \object{$\pi$\,Men\,c} a prototype for investigating alternative scenarios for the atmospheric composition of highly irradiated super-Earths, and its expected bulk density could represent a threshold which separates hydrogen-dominated from non hydrogen-dominated planets.

Another reason why \object{$\pi$\,Men} is very intriguing is that it hosts a Doppler-detected sub-stellar companion (minimum mass $m_{\rm b}\sin i_{\rm b}$ $\sim$10 $M_{\rm jup}$) on a long-period ($P_{\rm b}$ $\sim$2\,100 days) and very eccentric ($e_{\rm b}$ $\sim$0.6) orbit \citep{jones02}. This makes this system a nice laboratory to study the formation and dynamical evolution of planetary systems, with the benefit of accurate and precise planetary parameters determined with high-precision spectroscopy and transit photometry.

In this work, we present an updated characterization of the \object{$\pi$\,Men} system largely based on spectroscopic observations with ESPRESSO, new \textit{TESS} transit light curves of planet c, and astrometric data of the solar-type star from Hipparcos and Gaia. We revise the stellar and planetary fundamental parameters, unveiling the detailed 3-D system architecture for the first time.

%%%%%%%%%%%%%%%%%%%%%%%%%%%%%%%%%%%%%%%%%%%%%%%%%%%%%%%%%%%%%%%%%%%%%%%%%%%%%%%%%%%%%%%%%%%%%%%
\section{Overview of the new dataset}
The observations of \object{$\pi$\,Men} with ESPRESSO (using the instrument in single Unit Telescope mode with a median resolving power $R$=138\,000 over the 378.2 and 788.7\,nm wavelength range) were carried out within one of the sub-programmes of the Guaranteed Time Observations (GTO), aimed at using the very precise RVs to characterize (i.e. measuring masses and bulk densities) transiting planets discovered by \textit{TESS} and \textit{Kepler}'s second light \textit{K2} mission (see \citealt{Pepe-2020} for a detailed discussion of the ESPRESSO on-sky performance). \object{$\pi$\,Men} was observed starting from September 2018, right before the end of the commissioning phase of the instrument, up to March 2019. We collected 275 spectra in 37 nights (multiple and consecutive exposures per night) over a time span of 201 days. The spectra were acquired with a typical exposure time of 120\,s, providing median signal-to-noise ratio S/N=243 per extracted pixel at $\lambda=$\,500\,nm. 
In this work we also use previously unreleased spectra from CORALIE to extract additional RVs. \object{$\pi$\,Men} was observed with CORALIE from November 1998 to February 2020, collecting 60 spectra with a typical exposure time of 300-600 s (S/N=82-124 at 550 nm).

The spectroscopic follow-up with ESPRESSO further benefited of the simultaneous re-observations of \textit{TESS}, allowing for an improved synergy between spaced-based transit searches and ground-based radial-velocity observations. \textit{TESS} re-observed \object{$\pi$\,Men} during cycle 1 (sectors 4, 8, 11-13) from October 2018 to July 2019, gathering 19 additional transits of the planet in short-cadence mode.

%%%%%%%%%%%%%%%%%%%%%%%%%%%%%%%%%%%%%%%%%%%%%%%%%%%%%%%%%%%%%%%%%%%%%%%%%%%%%%%%%%%%%%%%%%%%%%%
\section{Stellar fundamental parameters and activity diagnostics from ESPRESSO spectra}
We obtained a combined ESPRESSO spectrum of \object{$\pi$\,Men} with a very high S/N>2\,000, and we analysed it to derive the basic stellar physical parameters summarized in Table\,\ref{tab:stellar_param}. A subset of blaze-corrected bi-dimensional (S2D) spectra at the barycentric reference frame were coadded, normalized, merged and corrected for RV (see Fig. \ref{fig:coaddedspectrum}) using the \textsc{StarII} workflow of the data analysis software (DAS) of ESPRESSO \citep{DiMarcantonio18}. The stellar parameters were derived using \textsc{ARES v2} and \textsc{MOOG2014} \citep[for more details, see][]{Sousa-2014} in which the spectral analysis is based on the excitation and ionization balance of the iron abundances. We used the ARES code \citep[][]{Sousa-2007, Sousa-2015} to consistently measure the equivalent widths for each line. The linelist used in this analysis was the same as in \citet[][]{Sousa-2008}. The abundances were computed in local thermodynamic equilibrium (LTE) with the MOOG code \citep[][]{Sneden-1973}. For this step a grid of plane-parallel Kurucz ATLAS9 model atmospheres was used\citep[][]{Kurucz-1993}. This is the same method used to derive homogeneous spectroscopic parameters for the Sweet-CAT catalogue \citep[][]{santos13, Sousa-2018}. The final uncertainties of the spectroscopic parameters are obtained from the formal errors by adding in quadrature 60\,K, 0.04\,dex, and 0.1\,dex for $T_{\rm eff}$, [Fe/H], and $\log g_{\rm \star}$, respectively, in order to take systematic errors into account, as described in \cite{sousa11}. Stellar mass, radius and age are derived using the optimization code PARAM \citep{dasilva06,rodrigues14,rodrigues17} with the additional information of \textit{Gaia} Data Release\,2 (DR2) parallax $\pi=54.705\pm0.067$\,mas and magnitude $G$= 5.4907$\pm$0.0014 mag \citep{gaia2018}, 
and 2MASS band $K_s=4.241\pm0.027$\,mag. These results are in agreement with those derived from HARPS spectra \citep{santos13}, with the age in agreement with that obtained by \cite{delgadomena15} and based on the lithium abundance determination.
We derived the projected rotational velocity $v\sini_{\star}$ using the package \textsc{FASMAsynthesis} \citep{tsantaki18}. We fixed the spectroscopic stellar parameters to the values derived by \textsc{ARES+MOOG}. The macroturbulance velocity in this analysis was set to 3.8 $\kms$ following the relation presented in \cite{doyle14}. The $v\sini_{\star}$ was the only free parameter used in this analysis where synthesis spectra are compared with our ESPRESSO combined spectrum for a bunch of FeI lines. We obtained $v\sini_{\star}$=3.34$\pm$0.07 \kms, larger than the value 2.96 \kms determined by \cite{delgadomena15} using HARPS spectra, and in agreement with the estimate of \cite{gandolfi18}.

\begin{table*}
\caption{Fundamental parameters of \object{$\pi$\,Men} derived from the analysis of ESPRESSO spectra. We also include the original values derived by \cite{gandolfi18} and \cite{huang18} for comparison. }            
\label{tab:stellar_param}      
\centering          
%\begin{small}
\begin{tabular}{l c c c }    
\hline\hline      
Parameter & This work & \cite{gandolfi18} & \cite{huang18} \\
\hline                    
   Effective temperature $T_{\rm eff}$ [K] & 5998$\pm$62 & 5870$\pm$50 & - \\  
   Surface gravity $\log g_{\rm \star}$ [cgs] & 4.43$\pm$0.10 & 4.36$\pm$0.02\tablefootmark{a} & - \\
                                              &               & 4.33$\pm$0.09\tablefootmark{b} &    \\   
   Iron abundance [Fe/H] [dex] & 0.09$\pm$0.04 & 0.05$\pm$0.09 & - \\
   Microturbulence $\xi$ [km s$^{-1}$] & 1.12$\pm$0.02 & - & - \\
   $v\sini_{\star}$ [\kms] & 3.34$\pm$0.07 & 3.3$\pm$0.5 & - \\
   Mass $M_{\rm \star}$ [$\Msun$] & 1.07$\pm$0.04 & 1.03$\pm$0.03 & 1.094$\pm$0.039 \\
   Radius $R_{\rm \star}$ [$\Rsun$] & 1.17$\pm$0.02 & 1.10$\pm$0.01 & 1.10$\pm$0.023  \\
   Density $\rho_{\rm \star}$ [$\rho_{\rm \odot}$] & 0.67$\pm$0.04 & - & 0.814$\pm$0.046\\ 
   Age [Ga] & 3.92$^{+1.03}_{-0.98}$ & 5.2$\pm$1.1 & 2.98$^{+1.4}_{-1.3}$ \\
\hline\hline                
\end{tabular}
\tablefoot{
\tablefoottext{a}{From spectroscopy and isochrones.}
\tablefoottext{b}{From spectroscopy.}
}
%\end{small}
\end{table*}

\begin{figure}
   \centering
   \includegraphics[width=\hsize]{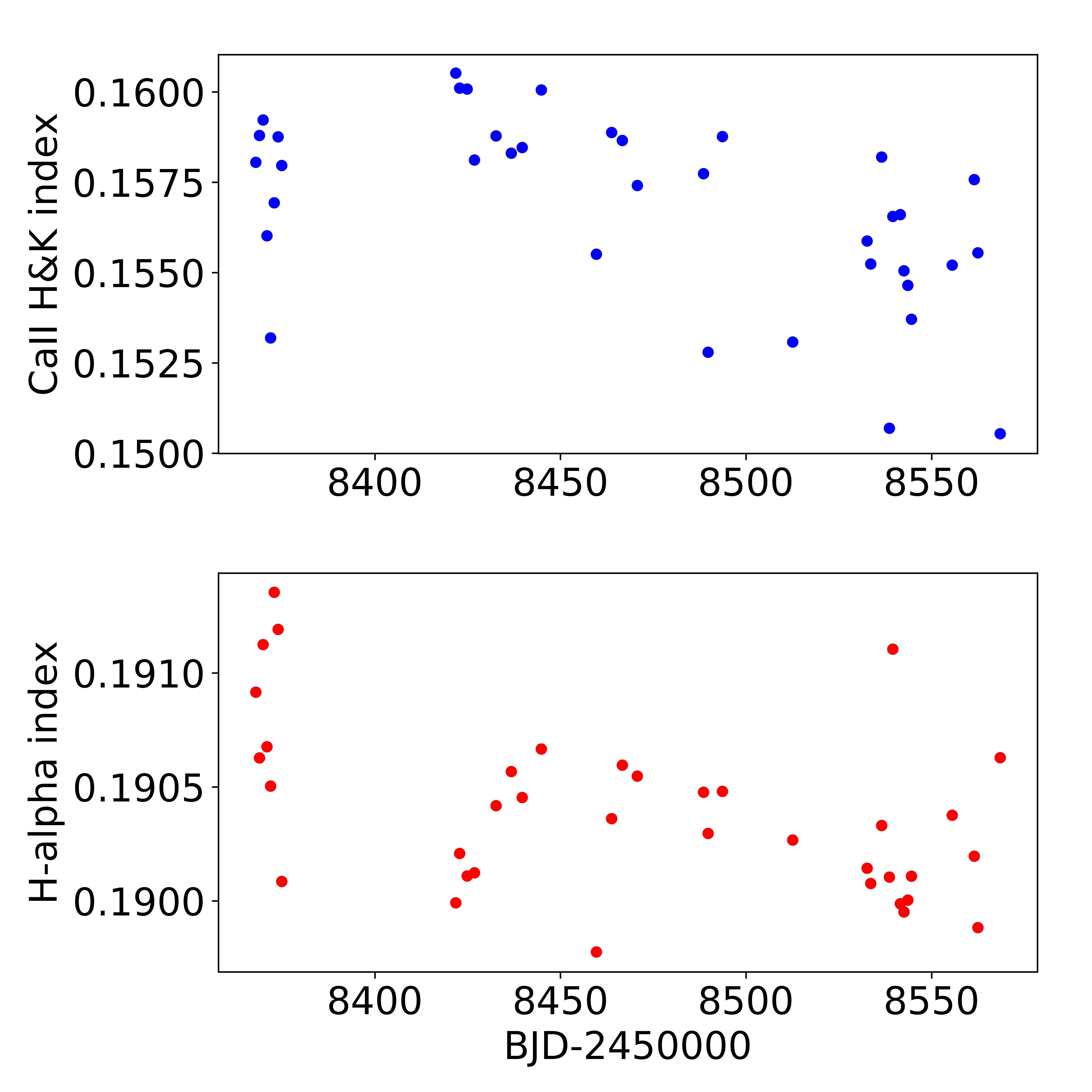}
      \caption{Time series of the S$_{\rm MW}$ (upper panel) and H-$\alpha$ (bottom panel) spectroscopic activity indexes derived from the ESPRESSO spectra.}
        \label{fig:smw_halpha_index}  
\end{figure}

The time series of the $S_{\rm MW}$ and H-$\alpha$ chromospheric activity indexes extracted from the ESPRESSO spectra are shown in Fig.\,\ref{fig:smw_halpha_index} (nightly averages). The $S_{\rm MW}$ index shows variations suggestive of a long-term cycle that we cannot characterize with the available data set. The Generalized Lomb-Scargle (GLS; \citealt{zech09}) periodograms of the H-$\alpha$ activity index and bisector asymmetry indicator $BIS$ of the cross-correlation function (CCF) show the main peak at the same period $P\sim$122\,days.

%%%%%%%%%%%%%%%%%%%%%%%%%%%%%%%%%%%%%%%%%%%%%%%%%%%%%%%%%%%%%%%%%%%%%%%%%%%%%%%%%%%%%%%%%%%%%%
\section{Radial velocities and photometry analysis}

\subsection{Data extraction} 

In this work we used RVs extracted from ESPRESSO spectra using the version 2.0.0 of the ESPRESSO data reduction pipeline\footnote{http://www.eso.org/sci/software/pipelines/} (DRS), adopting a template mask for a star of spectral type F9V to derive the CCF. During each observing night we collected series of multiple spectra at a rate of 2 to 12 consecutive exposures. Due to the technical intervention on ESPRESSO on September 2018 (close to the end of the commissioning phase), the RVs taken up to and after epoch BJD 2\,450\,8374 were treated in our analysis as two independent data sets composed of 71 and 204 measurements, respectively (reducing to 8 and 29 data points for nightly binned data), each characterized by an independent uncorrelated jitter and RV offset free parameter.

We also included RVs extracted from CORALIE spectra. During 21-year long scientific observations, CORALIE \citep{Udry-2000} underwent two significant upgrades in 2007 and 2014 which improved the RV precision. Both the intervention resulted in a small RV offset between the dataset \citep{Segransan-2020}, that we took into account in our analysis, also including distinct uncorrelated jitter terms. We refer to each dataset as CORALIE-98, CORALIE-07, CORALIE-14, for the period covering 1998-2007, 2007-2014 and 2014-now. CORALIE RVs are especially useful for further constraining the orbit of planet b.
ESPRESSO and CORALIE RVs are listed in Table\,\ref{data_pimensae} and \ref{data_pimensae_2}. We added to the new RVs those measured with HARPS and UCLES that are publicly available in \cite{gandolfi18}. A total of 520 RVs covering a time span of 8062 days were used in our study, and they are summarized in Table\,\ref{tab:rv_summary}.

Concerning new photometric data, we extracted and analysed the publicly available \textit{TESS} light curve to provide updated transit parameters. The data were downloaded from the  Mikulski Archive for Space Telescopes (MAST) portal\footnote{\url{mast.stsci.edu/portal/Mashup/Clients/Mast/Portal.html}}. For each sector, we de-trended the light curve with a spline filter and breakpoints every 0.5\,days to remove long-term stellar activity and instrumental trends similar to \citet{Barros2016}. Next, we extracted the region of the light curves with three times the transit duration and centred around the mid-transit times. Then, we fitted a first-order polynomial to the out-of-transit data of each transit to normalise it. We excluded the first transit of sector 1 (reference epoch BJD 2\,458\,325) and the first of sector 4 (reference epoch BJD 2\,458\,413) from our analysis since they are affected by instrumental systematics that cannot be corrected in a simple way. In particular, the second excluded transit is affected to such an extent that the transit shape is distorted and $\sim30\%$ deeper compared to the other transits. We analysed in total 22 transits (4 already published and 18 new).

\begin{table*}
\caption{Summary of the RVs analysed in this work. ESPRESSO data are distinguished by \textit{pre} and \textit{post} technical interventions, as described in the text. Values given for HARPS and ESPRESSO are for to unbinned data.}     \label{tab:rv_summary}      
\centering          
\begin{tabular}{c c c c c c}    
\hline\hline       
Instrument & Time span [BJD-2\,450\,000] & N$_{\rm meas.}$ & \multicolumn{2}{c}{Median $\sigma_{\rm RV}$ [m/s]} & Ref. \\
\hline
           &                             &                 & Binned & Unbinned & \\ 
\hline                    
   UCLES/AAT & 829.9-3669.2 & 42 & - & 4.6 & \cite{gandolfi18} \\  
   HARPS$_{\rm pre}$ & 3001.8-7033.6 & 128 & 0.30 & 0.50 & \cite{gandolfi18} \\
   HARPS$_{\rm post}$ & 7298.8-7464.5 & 16 & 0.25 & 0.40 & \cite{gandolfi18} \\
   CORALIE-98 & 1131.8-4108.7 & 10 & - & 5.32 & This work \\
   CORALIE-07 & 4433.7-6648.8 & 12 & - & 3.04 & This work \\
   CORALIE-14 & 7650.8-8891.5 & 38 & 2.76 & 2.76 & This work \\
   ESPRESSO$_{\rm pre}$ & 8367.8-8374.9 & 71 & 0.09 & 0.25 & This work \\
   ESPRESSO$_{\rm post}$ & 8421.8-8568.5 & 201 & 0.10 & 0.28 & This work \\
\hline\hline                  
\end{tabular}
\end{table*}

\subsection{Combined analysis}
\label{sec:comb_analysis}
We performed a combined light curve+RV fit using nightly binned RVs for HARPS, CORALIE and ESPRESSO, in order to average out short-term stellar jitter (e.g. p-modes, granulation). The analysis was carried out using the code presented in \cite{demangeon2018}. It combines parts of the Python packages \textsc{radvel} \citep{fulton2018} (\textsc{radvel.kepler.rv\_drive}) for the radial velocities and \textsc{batman} \citep{kreidberg2015} for the photometry into a Bayesian framework to compute the posterior probability of the planetary model parameters. This posterior function is then maximized using, first the Nelder-Mead simplex algorithm implemented in the Python package \texttt{scipy.optimize} \citep{gao2012} followed by an exploration with the affine-invariant ensemble sampler \textsc{mcmc} algorithm implemented by the Python package \textsc{emcee} \citep{foreman-mackey2013}.

As done by \cite{gandolfi18} and \cite{huang18}, we modeled any additional source of noise in the RVs not included in the nominal $\sigma_{\rm RV}(t)$ only by fitting uncorrelated jitter terms added in quadrature to $\sigma_{\rm RV}(t)$. In fact, we did not find evidence in the $TESS$ light curve, RVs, and spectroscopic activity diagnostics of a signal modulated over the stellar rotational period $\sim$18\,d found by \cite{zurlo18}, or its harmonics, such as to justify the use of a more sophisticated model to mitigate the stellar activity (e.g. Gaussian Process regression). To this regard, based on our measurements of $v\sini_{\star}$ and $R_{\rm \star}$, we estimated the upper limit of the rotation period of $\pi$ Men to be 17.7$\pm$0.5 d. We tested models with the eccentricity $e_c$ of \object{$\pi$ Men\,c} set to zero or fitted as free parameter, to explore if the use of ESPRESSO RVs and additional \textit{TESS} transits helps to constrain $e_{\rm c}$. This parameter wasn't well determined by \cite{huang18} and \cite{gandolfi18}, who could constrain $e_{\rm c}$ to be less than 0.3 and 0.45 (at 68$\%$ of confidence), respectively. 

\begin{figure}
   \centering
   \includegraphics[width=\hsize]{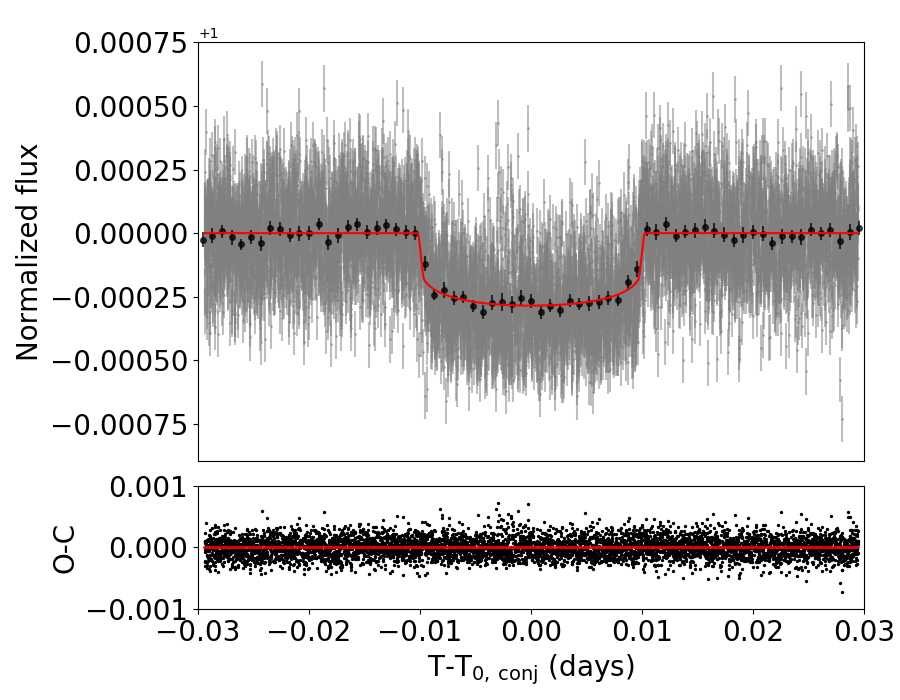}
      \caption{Transit signal of $\pi$\,Men\,c observed in the TESS light curve. Data from 22 individual transits are phase folded to the orbital period of the planet, and the red curve represents the best-fit model (Table\,\ref{tab:percentiles_rvmodel}).
              }
         \label{fig:bestfit_transit}
   \end{figure}

The results of our analysis and those from the literature are summarized in Table\,\ref{tab:percentiles_rvmodel}. Based on the Bayesian information criterion (BIC), we found that the model with fixed circular orbit is strongly favoured (BIC$_{\rm ecc}$-BIC$_{\rm circ}$>10), and we adopted it as our final solution. However, we cannot rule out a mild eccentricity (the posterior is not a zero-mean Gaussian distribution), and we are able to set the upper limit $e_{c}$<0.21 (corresponding to the 68-th percentile of the posterior), providing a further constraint for studies of the dynamical interaction with the massive and eccentric companion b. The best-fit model for the transit light curve of \object{$\pi$ Men\,c} (i.e. that obtained using the derived median values for the free parameters) is shown in Fig.\,\ref{fig:bestfit_transit}, with the 22 individual transits we analysed combined. Figure\,\ref{fig:spectorbits} shows the best-fit spectroscopic orbits for planets b and c. The mass of \object{$\pi$ Men\,c} $m_{\rm c}$=4.3$\pm0.7$ $\mearth$ is measured with a precision of $\sim$16$\%$. 

The upper panel of Fig.\,\ref{fig:residual_period} shows the GLS periodogram of the ESPRESSO RV residuals, after removing the signal induced by planet\,b. It clearly shows the dominant peak at the orbital period of \object{$\pi$\,Men\,c}, with a bootstrapping false alarm probability (FAP) of 0.6$\%$ determined from 10\,000 simulated datasets. The dispersion of the ESPRESSO residuals is 1.2\ms (middle panel of Figure\,\ref{fig:residual_period}), half the RMS of the HARPS residuals. 
The uncorrelated jitter term $\sigma_{\rm jitter,\:ESPRESSO}$ $\sim$1.2$\ms$ and the RMS of the residuals are one order of magnitude higher that the typical RV precision, and this could be due to effects of the stellar magnetic activity for which our model does not include an analytic term. We then considered the ESPRESSO residuals obtained after removing also the signal of \object{$\pi$\,Men\,c}, and we did not find significant correlations with the BIS, the $S_{\rm MW}$ and H-$\alpha$ activity diagnostics. Therefore, explaining the observed ``excess'' of jitter in terms of activity does not appear straightforward. The GLS periodogram of these residuals is shown in the last panel of Fig.\,\ref{fig:residual_period}. Through a bootstrap (with replacement) analysis, we found that the peak with the highest power has a false alarm probability (FAP) of $\sim$37$\%$. We further discuss this signal in Appendix \ref{appendixC}.

The ESPRESSO and CORALIE observations cover the periastron passage of planet b ($T_{\rm b,\:peri}=2458388.6\pm2.2$\,BJD), allowing for a more precise determination of the Doppler semi-amplitude $K_{\rm b}$ and eccentricity $e_{\rm b}$. With 18 additional transits of \object{$\pi$\,Men\,c} available, we improved the accuracy and precision of the transit ephemeris and of the transit depth, that in combination with the re-determined stellar radius resulted in a planet radius $R_{\rm c}=2.11\pm$0.05 $\rearth$, slightly larger than that reported in literature. 

\begin{figure}
   \centering
   \includegraphics[width=\hsize]{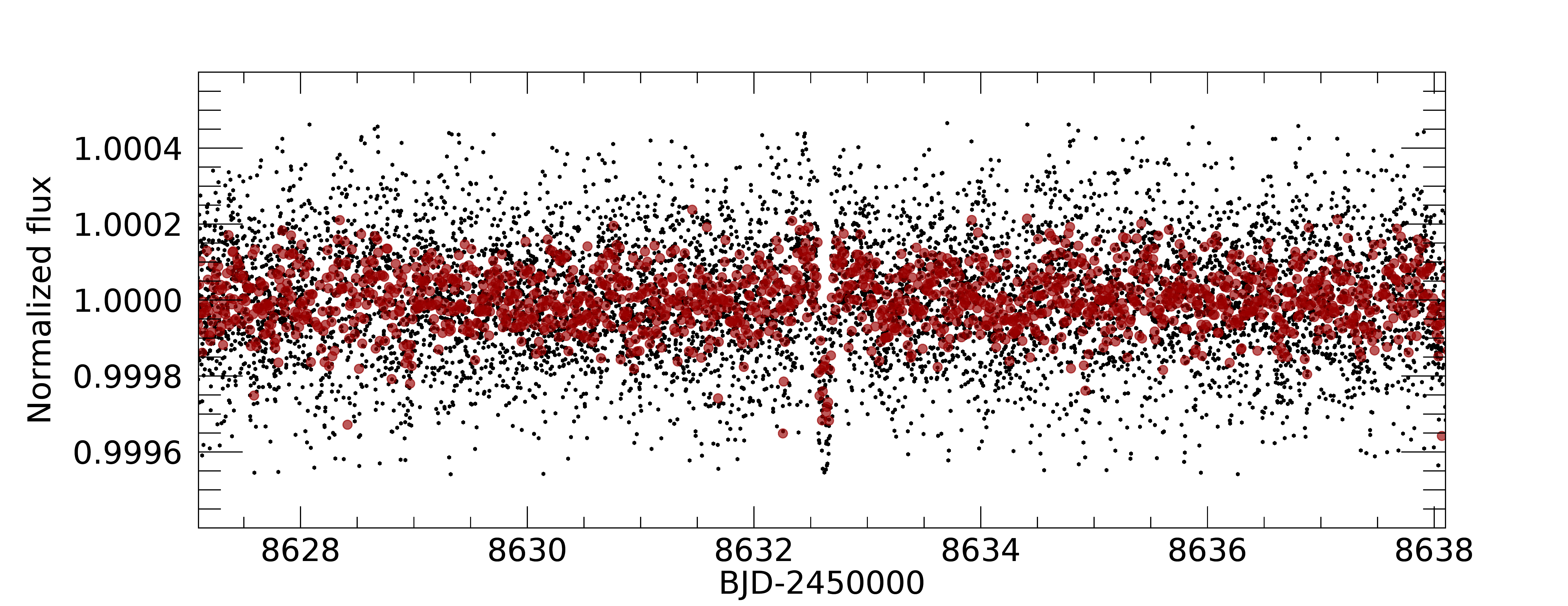}
      \caption{Portion of the \textit{TESS} light curve centered around the predicted time of inferior conjunction of $\pi$\,Men\,b ($\pm$5$\sigma$). Red dots represent averages of 5-data point bins. The only visible transit signal is that of $\pi$\,Men\,c.
              }
         \label{fig:transit_b}
   \end{figure}

According to our results, the predicted time of inferior conjunction $T_{\rm b,\:conj}$ of the outermost companion falls within the time span of the \textit{TESS} observations (sector 12). We checked the light curve within a $\pm$5$\sigma$ range from the best-fit value (Fig.\,\ref{fig:transit_b}), and we did not find evidence for the transit of \object{$\pi$\,Men\,b}. Assuming a radius of 0.8 $\Rjup$ for \object{$\pi$\,Men\,b} \citep{sorahana13}, we could detect transits of the sub-stellar companion if the orbital inclination angle $i_{\rm b}$ were within the \textit{penumbra} cone defined by the narrow angle $\pm$ $\sim$0.1 deg as measured from a perfectly edge-on orbit. 

\begin{table*}
       \caption{Best-fit results of the $\pi$\,Men photometry+RV joint modelling. Values are given as the $50^{\rm th}$ percentile of the posterior distributions, and the uncertainties are derived from the $16^{\rm th}$ and $84^{\rm th}$ percentiles.} \label{tab:percentiles_rvmodel}
\begin{tiny} 
\begin{tabular}{lcccc}
    %     \scriptsize
    %    \centering
           \hline\hline
            \noalign{\smallskip}
           Jump parameter &  Prior & \multicolumn{2}{c}{Best-fit value} & Literature \\
           \noalign{\smallskip}
             \noalign{\smallskip}
                          &        & $\pi$ Men\,c: circular (adopted) & $\pi$ Men\,c: eccentric & \\
             \noalign{\smallskip}
    %         \hline
    %        \noalign{\smallskip}
            \hline
            \noalign{\smallskip}
            $K_{\rm b}$ [m$\,s^{-1}$] & $\mathcal{U}$(185.6,199.6) & 196.1$\pm$0.7 & 196.0$_{-0.6}^{+0.7}$ & 195.8$\pm$1.4 (a); 192.6$\pm$1.4 (b) \\
            \noalign{\smallskip}
            $P_{\rm b}$ [days] & $\mathcal{U}$(2084.42, 2101.72) & 2088.8$\pm$0.4 & 2088.8$\pm$0.4 & 2091.2$\pm$2.0 (a); 2093.07$\pm$1.73 (b) \\ 
            \noalign{\smallskip}
            $T_{\rm b,\:conj}$ [BJD-2\,450\,000] & $\mathcal{U}$(8590.2, 8674.2) & 8632.6$\pm$1.1 & 8632.7$\pm$1.2 & 6548.2$\pm$2.7 (a); -3913.0$\pm$8.4 (b) \\
            \noalign{\smallskip}
            $e_{\rm b}\cos\omega_{\rm \star,\: b}$ & $\mathcal{U}$ (-1, 1) & 0.5552$\pm$0.0014  & 0.5551$_{-0.0015}^{+0.0014}$ & - \\
            \noalign{\smallskip}
            $e_{\rm b}\sin\omega_{\rm \star,\: b}$ & $\mathcal{U}$ (-1, 1) & -0.3220$_{-0.0028}^{+0.0027}$ & -0.3221$_{-0.0029}^{+0.0028}$ & - \\
            \noalign{\smallskip}
            $K_{\rm c}$ [m$\,s^{-1}$] & $\mathcal{U}$(0,5) & 1.5$\pm$0.2 & 1.5$_{-0.2}^{+0.3}$ & 1.55$\pm$0.27 (a); 1.58$^{+0.26}_{-0.28}$ (b) \\ 
            \noalign{\smallskip}
            $P_{\rm c}$ [days] & $\mathcal{N}$(6.2679,0.00046) & 6.267852$\pm$0.000016 & 6.267852$^{+0.000017}_{-0.000016}$ & 6.26834$\pm$0.00024 (a); 6.2679$\pm$0.00046 (b) \\ 
            \noalign{\smallskip}
            $T_{\rm c,\:conj}$ [BJD-2\,450\,000] & $\mathcal{N}$(8519.8066,0.0012) & 8519.8068$\pm$0.0003 & 8519.8065$_{-0.0006}^{+0.0005}$ & 8325.503055$\pm$0.00077 (a); 8325.50400$^{+0.0012}_{-0.00074}$ (b) \\ 
            \noalign{\smallskip}
            $e_{\rm c}\cos\omega_{\rm \star,\: c}$ & $\mathcal{U}$ (-1, 1) (c) & 0 (fixed) & -0.03$\pm$0.06 &  0 (a,b) \\
            \noalign{\smallskip}
            $e_{\rm c}\sin\omega_{\rm \star,\: c}$ & $\mathcal{U}$ (-1, 1) (c) & 0 (fixed) & -0.11$_{-0.17}^{+0.16}$ &  0 (a,b) \\
            \noalign{\smallskip}
    %        \hline
    %        \noalign{\smallskip}
    %        \hline
   %         \noalign{\smallskip}
            $\sigma_{\rm jit, UCLES}$ [m$\,s^{-1}$] & $\mathcal{U}$(0,50) & 4.1$_{-0.9}^{+1.0}$ & 4.0$^{+1.0}_{-0.9}$ & 4.26$^{+1.10}_{-0.96}$ (a); 6.7$\pm$0.60 (b) \\ 
            \noalign{\smallskip}
            $\sigma_{\rm jit, CORALIE-98}$ [m$\,s^{-1}$] & $\mathcal{U}$(0,50) & 4.3$_{-0.9}^{+1.0}$ & 4.3$\pm$0.1 & - \\ 
            \noalign{\smallskip}
             $\sigma_{\rm jit, CORALIE-07}$ [m$\,s^{-1}$] & $\mathcal{U}$(0,50) & 13.3$_{-3.3}^{+4.9}$ & 13.7$^{+5.2}_{-3.4}$ & - \\ 
            \noalign{\smallskip}
             $\sigma_{\rm jit, CORALIE-14}$ [m$\,s^{-1}$] & $\mathcal{U}$(0,50) & 13.2$_{-2.7}^{+3.7}$ & 13.4$^{+4.3}_{-2.7}$ & - \\ 
            \noalign{\smallskip}
            $\sigma_{\rm jit, HARPS_{pre-upgrade}}$ [m$\,s^{-1}$] & $\mathcal{U}$(0,10) & 2.3$\pm$0.3 & 2.3$\pm$0.3 & 2.35$^{+0.19}_{-0.17}$ (a); 2.33$\pm$0.18 (b) \\ 
            \noalign{\smallskip}
            $\sigma_{\rm jit, HARPS_{post-upgrade}}$ [m$\,s^{-1}$] & $\mathcal{U}$(0,10) & 1.8$_{-0.4}^{+0.6}$ & 1.8$_{-0.4}^{+0.6}$ & 1.69$^{+0.39}_{-0.29}$ (a); 1.74$\pm$0.33 (b) \\ % 
            \noalign{\smallskip}
            $\sigma_{\rm jit, ESPRESSO_{pre-interv.}}$ [m$\,s^{-1}$] & $\mathcal{U}$(0,10) & 1.2$_{-0.3}^{+0.4}$ & 1.2$_{-0.3}^{+0.5}$ & - \\ 
            \noalign{\smallskip}
            $\sigma_{\rm jit, ESPRESSO_{post-interv.}}$ [m$\,s^{-1}$] & $\mathcal{U}$(0,10) & 1.2$\pm$0.2 & 1.3$\pm$0.3 & - \\ 
            \noalign{\smallskip}
            $\gamma_{\rm CORALIE-98}$ [m$\,s^{-1}$] & $\mathcal{U}$(10600,10800) (d) & 10674.0$_{-4.8}^{+4.6}$ & 10674.6$\pm$5.0 & - \\
             \noalign{\smallskip}
            $\gamma_{\rm CORALIE-07}$ [m$\,s^{-1}$] & $\mathcal{U}$(-100,+100) (e) & -3.2$_{-6.1}^{+6.4}$ & -3.6$_{-6.5}^{+6.7}$ & - \\
             \noalign{\smallskip}
            $\gamma_{\rm CORALIE-14}$ [m$\,s^{-1}$] & $\mathcal{U}$(0,200) (e) & 21.9$_{-4.6}^{+4.8}$ & 21.5$\pm$5.0 & - \\
            \noalign{\smallskip}
            $\gamma_{\rm HARPS_{pre-upgrade}}$ [m$\,s^{-1}$] & $\mathcal{U}$(10600,10800) (d) & 10707.0$\pm$1.0 & 10707.0$\pm$1.1 & - \\
            \noalign{\smallskip}
            $\gamma_{\rm HARPS_{post-upgrade}}$ [m$\,s^{-1}$] & $\mathcal{U}$(-10,+40) (e) & 22.7$\pm$0.8 & 22.7$\pm$0.8 & - \\
            \noalign{\smallskip}    
            $\gamma_{\rm ESPRESSO_{pre-interv.}}$ [m$\,s^{-1}$] & $\mathcal{U}$(10600,10800) (d) & 10639.0$\pm$2.0 & 10639.1$\pm$2.0 & -  \\ 
            \noalign{\smallskip}
            $\gamma_{\rm ESPRESSO_{post-interv.}}$ [m$\,s^{-1}$] & $\mathcal{U}$(-30,+10) (e) & -1.3$\pm$2.0 & -1.3$\pm$2.0 & - \\ 
            \noalign{\smallskip}
     %       \hline
    %        \noalign{\smallskip}
     %       \textbf{Transit parameters} & \\ 
            \noalign{\smallskip}
            $R_{\rm c}/R_{*}$ & $\mathcal{U}$(0,0.1) & 0.0165$\pm$0.0001 & 0.0166$\pm$0.0004 & 0.01721$\pm$0.00024 (a); 0.01703$_{-0.00023}^{+0.00025}$ (b) \\
            \noalign{\smallskip}
            $a_{\rm c}/R_{*}$ & (f) & 12.5$\pm$0.3 & 11.2$\pm$1.9 & 13.10$\pm$0.18 (a); 13.38$\pm$0.26 (b) \\
            \noalign{\smallskip}
            $i_{\rm c}$ [deg] & $\mathcal{U}$(0,90) (g) & 87.05$\pm$0.15 & 86.9$^{+0.6}_{-0.4}$ & 87.31$\pm$0.11 (a); 87.456$_{-0.076}^{+0.085}$ (b)\\
            \noalign{\smallskip}
            $\sigma_{\rm jit, TESS}$ [ppm] & $\mathcal{U}$(0,300) & 130$\pm$2 & 130$\pm$2 & - \\
            \noalign{\smallskip}
            limb darkening coeff $q_1$ & $\mathcal{N}$(0.280,0.002) & 0.280$\pm$0.002 & 0.280$\pm$0.002 & \\
            \noalign{\smallskip}
            limb darkening coeff $q_2$ & $\mathcal{N}$(0.270,0.002) & 0.270$\pm$0.002 & 0.270$\pm$0.002 & \\
            \noalign{\smallskip}
            \textbf{Derived planetary parameters} && \\
    %        \noalign{\smallskip}
    %        \hline
             \noalign{\smallskip}
             Eccentricity, $e_{\rm b}$ & & 0.642$\pm$0.001 & 0.642$\pm$0.001 & 0.6394$\pm$0.0025 (a); 0.637$\pm$0.002 (b) \\  
            \noalign{\smallskip}
             Argument of periastron, $\omega_{\rm \star,\: b}$ [deg] & & -30.1$\pm$0.3 & -30.1$\pm$0.3 & -29.3$\pm$0.7 (a); -29.4$\pm$0.3 (b) \\ 
             \noalign{\smallskip}
             $T_{\rm b,\:periastron}$ [BJD-2\,450\,000] & & 8388.6$\pm$2.2 & 8387.4$\pm$2.2 & \\
             \noalign{\smallskip}
             Minimum mass, $m_{\rm b}\sin{i_{\rm b}}$ [$\Mjup$] & & 9.89$\pm$0.25 & 9.89$\pm$0.25 & 9.66$\pm$0.20 (a); 10.02$\pm$0.15 (b) \\
             \noalign{\smallskip}
             Orbital semi-major axis, $a_{\rm b}$ [au] & & 3.28$\pm$0.04 & 3.28$\pm$0.04 & 3.22$\pm$0.03 (a); 3.10$\pm$0.02 (b)\\
             \noalign{\smallskip}
             Eccentricity, $e_{\rm c}$ & & 0 (fixed) & 0.15$^{+0.14}_{-0.08}$ (<0.21 68$\%$) & 0 (a,b) \\
             \noalign{\smallskip}
             Argument of periastron, $\omega_{\rm \star,\: c}$ [deg] & & 90 (fixed) & -93.7$_{-25.5}^{+182.8}$ & 0 (a,b) \\ 
             \noalign{\smallskip}
             Orbital semi-major axis, $a_{\rm c}$ [au] & & 0.0680$\pm$0.0008 & 0.0680$\pm$0.0009  & 0.06702$\pm$0.00109 (a); 0.06839$\pm$0.00050 (b) \\
             \noalign{\smallskip}
             Mass, $m_{\rm c}$ [$\mearth$] & & 4.3$\pm$0.7 & 4.5$\pm$0.7 & 4.52$\pm$0.81 (a); 4.82$^{+0.84}_{-0.86}$ (b) \\
             \noalign{\smallskip}
             Radius, $R_{\rm c}$ [\rearth] & & 2.11$\pm$0.05 & 2.11$\pm$0.07 & 2.06$\pm$0.03 (a); 2.04$\pm$0.05 (b) \\
             \noalign{\smallskip}
             Average density, $\rho_{\rm c}$ [$g$ $cm^{\rm -3}$] & & 2.8$\pm$0.5 & 2.8$\pm$0.5 & 2.82$\pm$0.53 (a); 2.97$^{+0.57}_{-0.55}$ (b) \\
             \noalign{\smallskip}
             \hline
             \noalign{\smallskip}
             $\Delta$BIC        & & 0 & +19 \\
             \noalign{\smallskip}
            RMS of the RV residuals [\ms] & All data & 5.6 &  \\
                                & HARPS  & 2.2 \\
                                & ESPRESSO  & 1.2  \\ 
            \noalign{\smallskip}
            \hline\hline
\end{tabular}   
\tablefoot{(a) After \cite{gandolfi18}. (b) After \cite{huang18}. (c) The eccentricity was further constrained to values $<$\,0.75. (d) Relative to the UCLES dataset, which is used as reference. (e) Relative to the \textit{pre-} dataset of the corresponding instrument.
(f) In the analysis we used the stellar density $\rho_{*}$ [$\rho_{\odot}$] as free parameter ($\mathcal{N}$(0.67,0.04)), from which we derived $a_{\rm c}/R_{*}$ at each step of the MC sampling.
(g) We used $\cos i_{\rm c}$ as free parameter.
}
\end{tiny}    
\end{table*}

\subsection{Mass limits for co-orbital companions to $\pi$ Men\,c}
\label{subsec:coorbitals}

Given the high-precision of the ESPRESSO and HARPS RVS, we explored the possibility of the presence of co-orbital bodies to \object{$\pi$ Men\,c} through the technique described in \cite{leleu17} and subsequently applied by \cite{lillo-box18a,lillo-box18b}. This technique uses the information from the transit time of the planet and the full radial velocity dataset to constrain the time lag between the planet transit and the time of zero radial velocity (a generalization of the \citealt{ford06} methodology). The technique is based on modeling the radial velocity data by using Eq.~18 from \cite{leleu17}, where the key parameter $\alpha$ contains all the information about the co-orbital signal. A posterior distribution of $\alpha$ compatible with zero discards the presence of co-orbitals up to a certain mass. Contrarily, if $\alpha$ is significantly different from zero, the data contains hints for the presence of a co-orbital body, with negative values corresponding to L$_{4}$ and positive values to L$_{5}$. In this framework, we analyse the radial velocity dataset with such model and using the \textsc{emcee} Markov Chain Monte Carlo ensemble sampler to explore the parameter space. We used the same priors for the parameters in common with the analysis presented in Sect. \ref{sec:comb_analysis} and a uniform prior $\mathcal{U}(-1,1)$ for the $\alpha$ parameter. We used 96 random walkers and 10\,000 steps per walker with two first burn-in chains and a final production chain with 5\,000 steps. We checked the convergence of the chains by estimating the autocorrelation times and checking that the chain length is at least 30 times this autocorrelation time for all parameters. The result provides a value for the $\alpha$ parameter of $\alpha=-0.25^{+0.19}_{-0.21}$. Although shifted towards negative values, the posterior distribution is compatible with zero at the 95$\%$ confidence level. Consequently, we cannot confirm the presence of co-orbitals. However, given this posterior value, we can certainly put upper limits to the presence of co-orbitals at both Lagrangian points. By using the 95$\%$ confidence levels, we can discard co-orbitals more massive than $3.1~M_{\oplus}$ at L$_{4}$ and $0.3~M_{\oplus}$ at L$_{5}$, i.e. co-orbitals more massive than three times the mass of Mars at L$_{5}$. An intensive and dedicated effort with additional radial velocity data would then be needed to further explore the L$_{4}$ region.

%%%%%%%%%%%%%%%%%%%%%%%%%%%%%%%%%%%%%%%%%%%%%%%%%%%%%%%%%%%%%%%%%%%%%%%%%%%%%%%%%%%%%%%%%%%

\section{Constraining the relative alignment of the planetary orbital planes}
\label{sec:astrometry}
We further constrain the relative alignment of the orbital planes of the two planets using the combination of the high-precision spectroscopic orbit for \object{$\pi$\,Men\,b} obtained 
thanks to the contribution of the ESPRESSO data set, and the absolute astrometry of Hipparcos and Gaia, as follows.

We first take the cross-calibrated \textit{Hipparcos/Gaia} DR2 $\pi$ Men proper motion values and the scaled Hipparcos-Gaia positional difference from the \cite{brandt18,brandt19b}\footnote{The original catalogue presented in
\cite{brandt18} is superseded by the new version published in \cite{brandt19b} that corrects an error in the calculation of the perspective
acceleration in R.A.} catalogue of astrometric accelerations. The latter quantity is defined as the difference in astrometric position between the two catalogues divided by the 
corresponding $\sim$25-yr time baseline, a factor of $\sim$4.4 longer than the orbital period of \object{$\pi$\,Men\,b}. It corresponds to a long-term proper-motion vector that can be considered as a close representation of the tangential velocity of the barycentre of the system. By subtracting this long-term proper motion from the quasi-instantaneous proper motions of the two catalogues one obtains a pair of 'proper motion difference', 'astrometric acceleration', or 'proper motion anomaly' 
values, in short $\Delta\mu$, assumed to be entirely describing the projected velocity of the photocenter around the barycentre at the Hipparcos and Gaia DR2 epochs\footnote{For recent applications of this technique for the detection of stellar and substellar companions see e.g. \citealt{calissendorff18,snellenbrown18, 
kervella19,brandt19,dupuy19,feng19,grandjean19}.}. 
The observed $\Delta\mu$ values (see Table\,\ref{tab:pimen_astro}) contain information on the orbital motion of \object{$\pi$\,Men\,b} (the orbital effect due to \object{$\pi$\,Men\,c} is entirely negligible). 
We elect to use the $\pi$ Men proper motion vector from the \cite{brandt18,brandt19b} catalogue instead of the physically equivalent and equally well validated quantity in the catalogue produced by \cite{kervella19} for two reasons: a) the former catalogue is constructed based on a linear combination of the two Hipparcos reductions, a choice that appears preferable with respect to considering either reduction individually; b) \cite{brandt18,brandt19b} brings the composite Hipparcos astrometry on the bright reference frame of Gaia DR2, resulting in an updated error model with rather conservative uncertainties, which are shown to be statistically well-behaved. The robustness of the \cite{brandt18,brandt19b} catalogue has been further probed recently by \cite{lindegren20,lindegren20b} when comparing the spin and orientation of the bright reference frame of Gaia DR2 using very long baseline interferometry observations of radio stars and the independent assessment of the rotation made by \cite{brandt18,brandt19b}.

We then follow \cite{kervella20} and explore via an MCMC algorithm the ranges of inclination $i_b$ and longitude of the ascending node $\Omega_b$ compatible with the 
absolute astrometry and the spectroscopically determined orbital parameters (and their uncertainties). The values of $i_b$ and $\Omega_b$ (using uniform priors on $\cos i_b$ and $\Omega_b$ over the allowed ranges for both prograde and retrograde motion) are fitted in a model of the proper motion differences that we build averaging over the actual \textit{Hipparcos} and \textit{Gaia} observing windows, adopting the times of \textit{Hipparcos} observations 
available in the \textit{Hipparcos}-2 catalogue \citep{vanleeu2007} and taking the Gaia transit times from the \textit{Gaia} Observation Forecast Tool (GOST)\footnote{\url{https://gaia.esac.esa.int/gost/index.jsp}}. This allows us to cope with the 'smearing' effect of the orbital motion due to the fact that the observed $\Delta\mu$ values are time averages of the intrinsic velocity vector of the star over the \textit{Hipparcos} and \textit{Gaia} observing periods, respectively. For \object{$\pi$\,Men\,b} this effect in non-negligible (see \citealt{kervella19}). 

\begin{figure}
   \centering
   \includegraphics[width=\hsize]{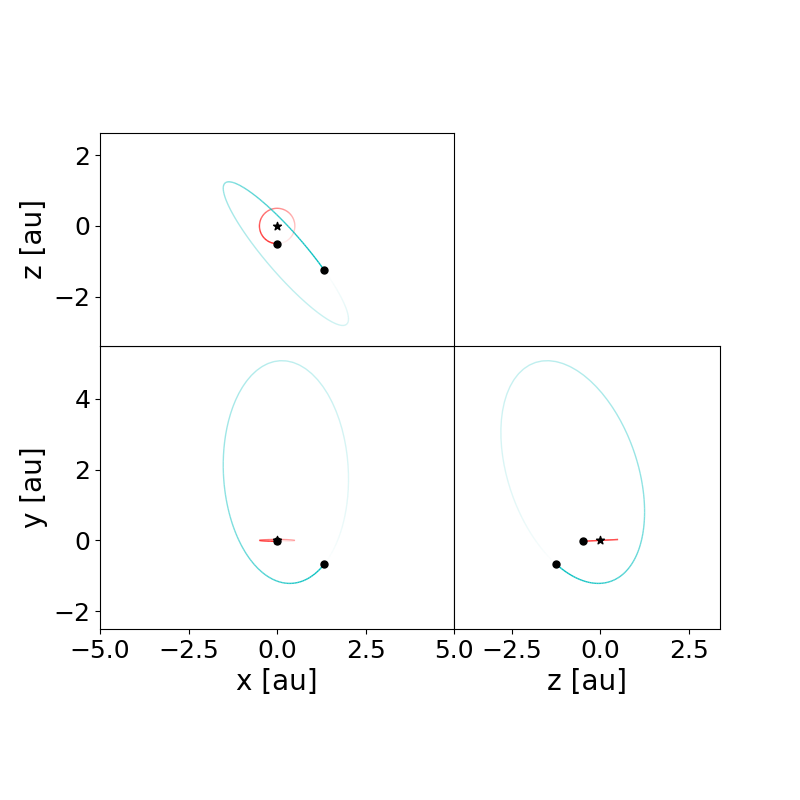} \\
      \caption{Sketch of the planets' orbits in Jacobi coordinates, showing the 3-D architecture of the \object{$\pi$\,Men} system. The orbits are not to scale, in that that of planet c has been enlarged to better show the mutual inclinations between the planetary orbital planes.}
         \label{fig:3d_orbits}
\end{figure}

The orbital fit results to the \textit{Hipparcos/Gaia} absolute astrometry are reported in Table\,\ref{tab:pimen_astro}, while in Figure\,\ref{fig:posteriors_astro} we show the posterior distributions for the model parameters explored in our MCMC analysis. The corresponding inferred true mass of \object{$\pi$\,Men\,b} is $m_b$=14.1$_{-0.4}^{+0.5}$\,M$_\mathrm {Jup}$. 
Furthermore, the evidence for orbital motion at both the \textit{Hipparcos} and \textit{Gaia} epochs also allows us
to break the degeneracy between prograde and retrograde motion, the latter being clearly favoured. Overall, the inference is for a highly significant non-coplanarity 
between $\pi$\,Men\,b and $\pi$\,Men c. Given that the inclination of the orbital plane of the latter is known, we can then directly provide constraints on the possible 
range of mutual inclination angles $i_\mathrm{rel}$, expressed as a function of the unknown longitude of the ascending node of \object{$\pi$\,Men\,c} (allowed to vary in the range [0,360]\,deg). The results are shown 
in Figure\,\ref{fig:rel_inc}. We find that 52.3$\leq i_\mathrm{rel}\leq$ 128.8\,deg, at the $1\sigma$-level. The $i_\mathrm{rel}$ distribution shows two clear peaks 
at 50\,deg and 130\,(=180-50)\,deg. A sketch of the 3-D system's architecture is given in Figure \ref{fig:3d_orbits}.

%\begin{figure}
%   \centering
%   \includegraphics[width=\hsize]{corner_ecc_omega.pdf}
%      \caption{Posterior distributions for $e_{\rm c}$ and $\omega_{\rm \star,\:c}$ from the MCMC analysis.}
%        \label{fig:ecc_omega_post}  
%\end{figure}

\begin{figure}
   \centering
   \includegraphics[width=\hsize]{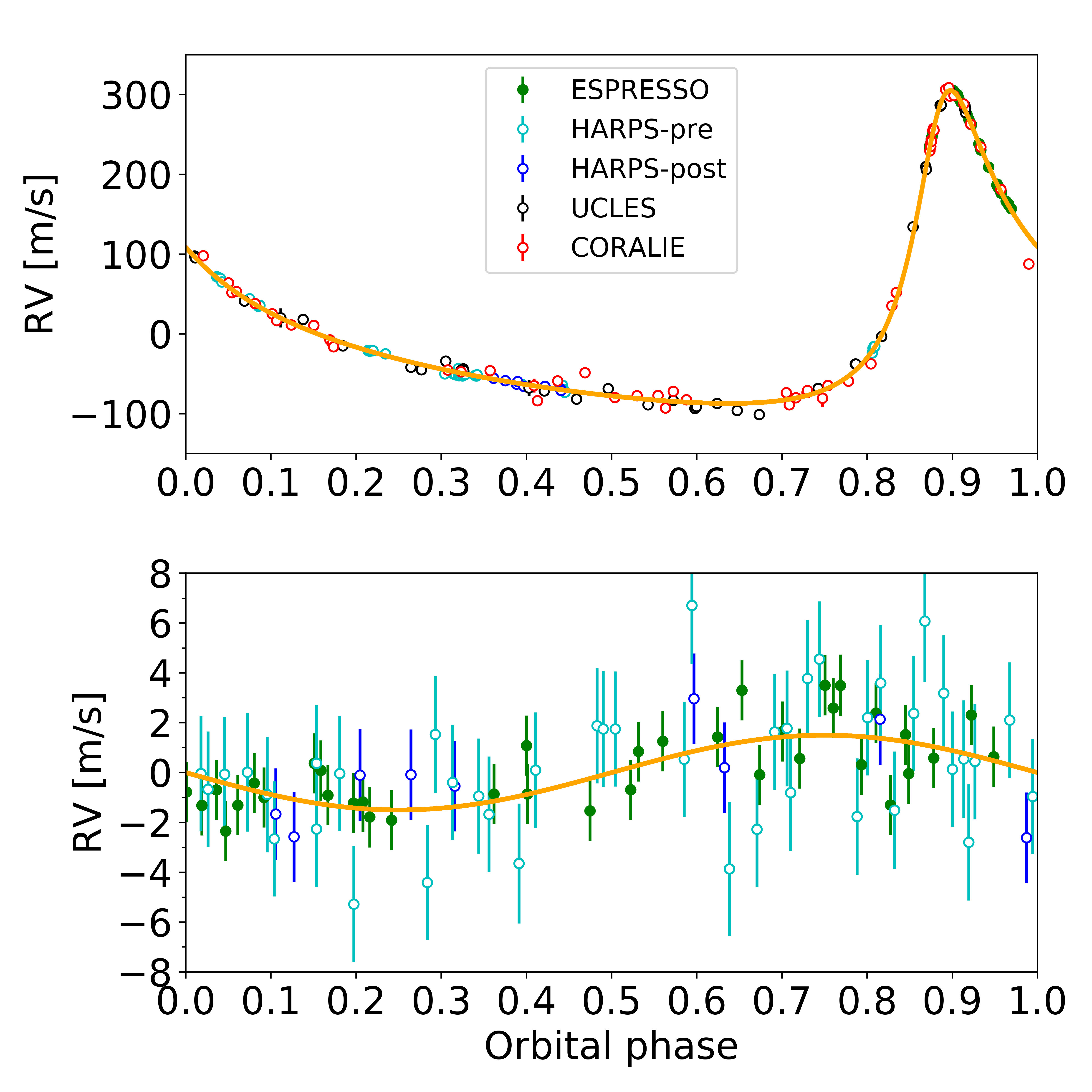}
      \caption{Spectroscopic orbits of the two planets in the $\pi$\,Men system (upper panel: \object{$\pi$\,Men\,b}; lower panel: \object{$\pi$\,Men\,c}; best-fit solutions in Table\,\ref{tab:percentiles_rvmodel}). The orange curve represents the best-fit model. For $\pi$ Men\,c we do not show the more scattered and less precise UCLES and CORALIE data for a better visualization, and the error bars include uncorrelated jitters added in quadrature to $\sigma_{\rm RV}$.
              }
         \label{fig:spectorbits}
   \end{figure}
   
\begin{figure}
   \centering
   \includegraphics[width=7cm]{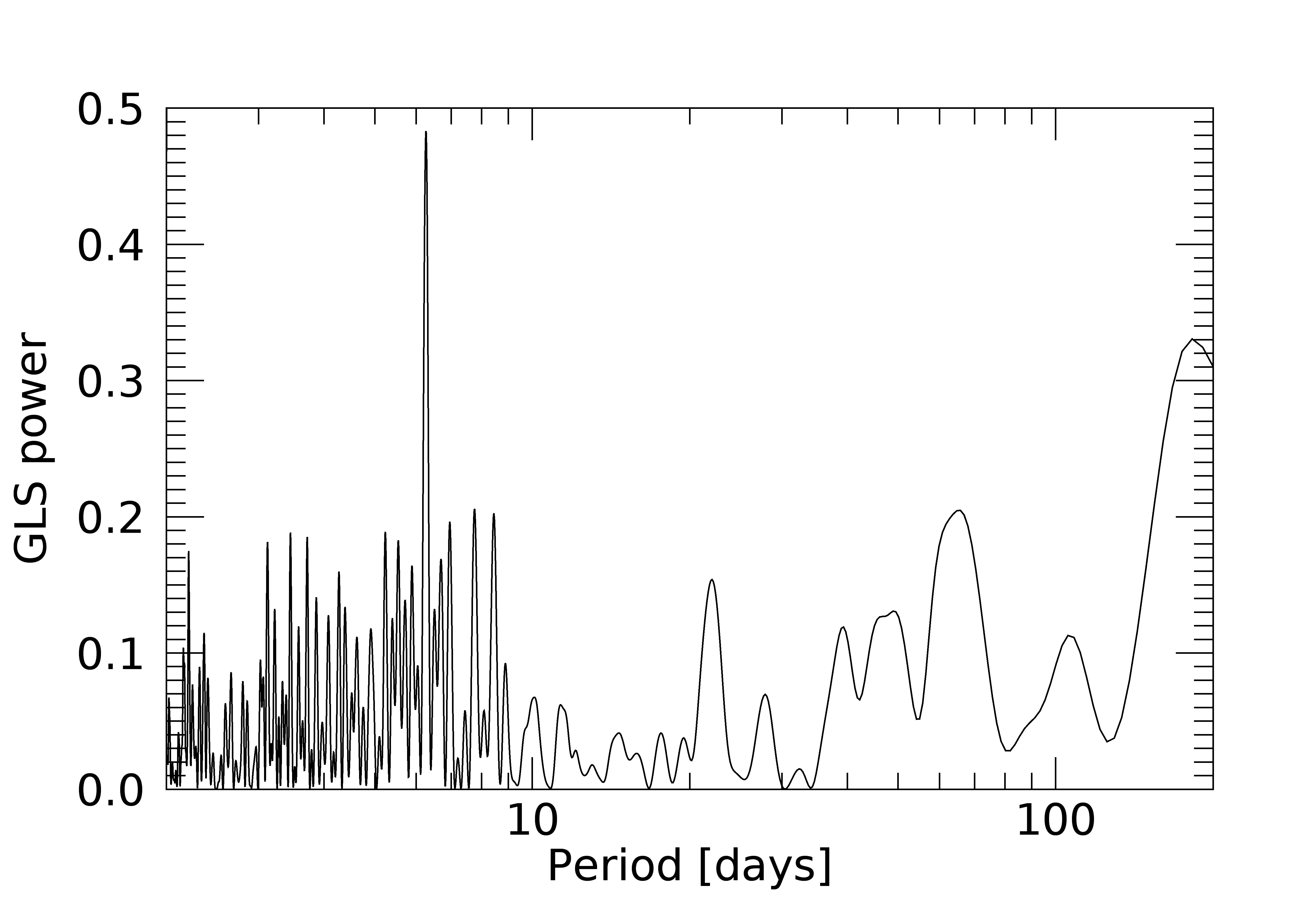}
   \includegraphics[width=7cm]{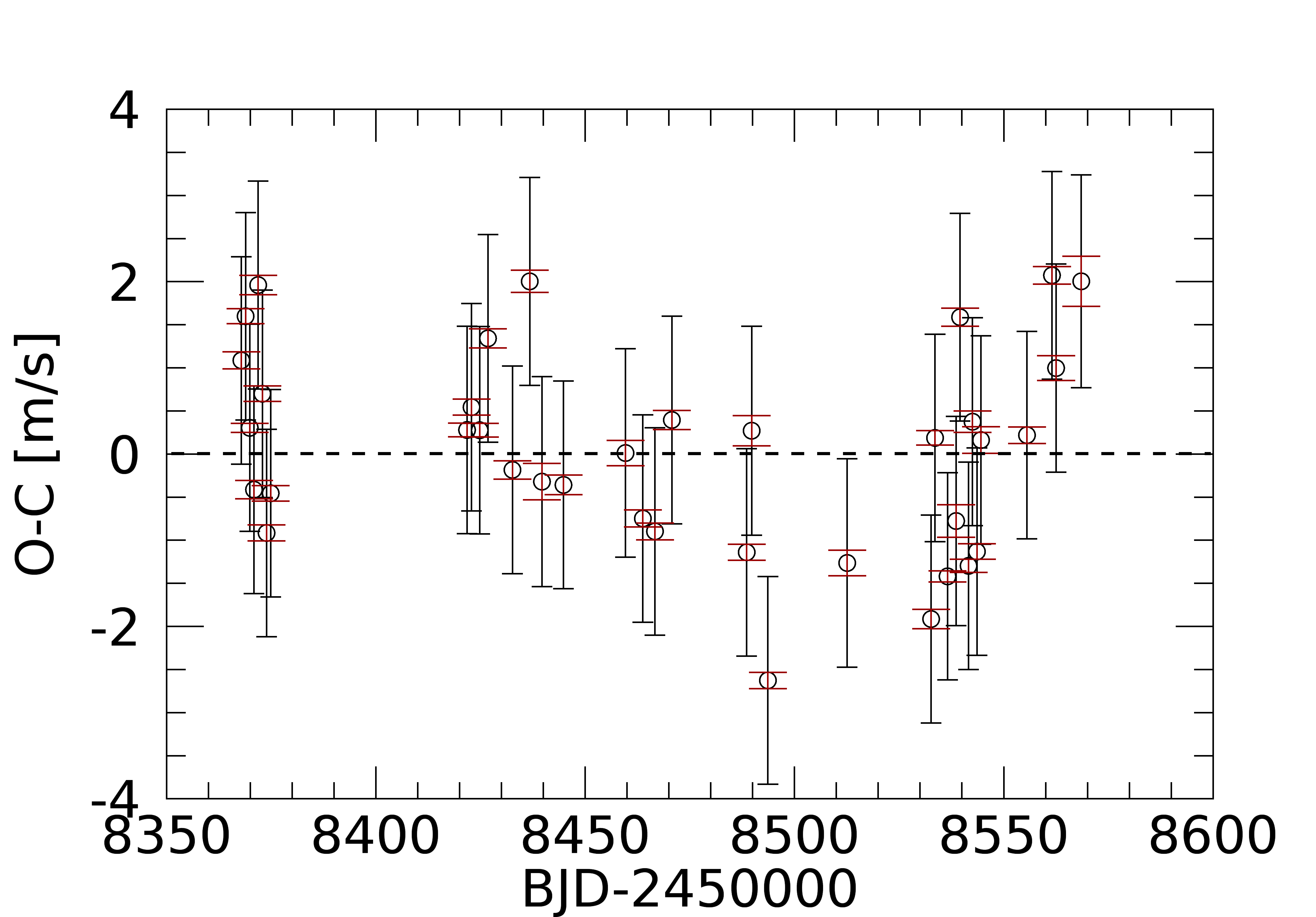}
  \includegraphics[width=7cm]{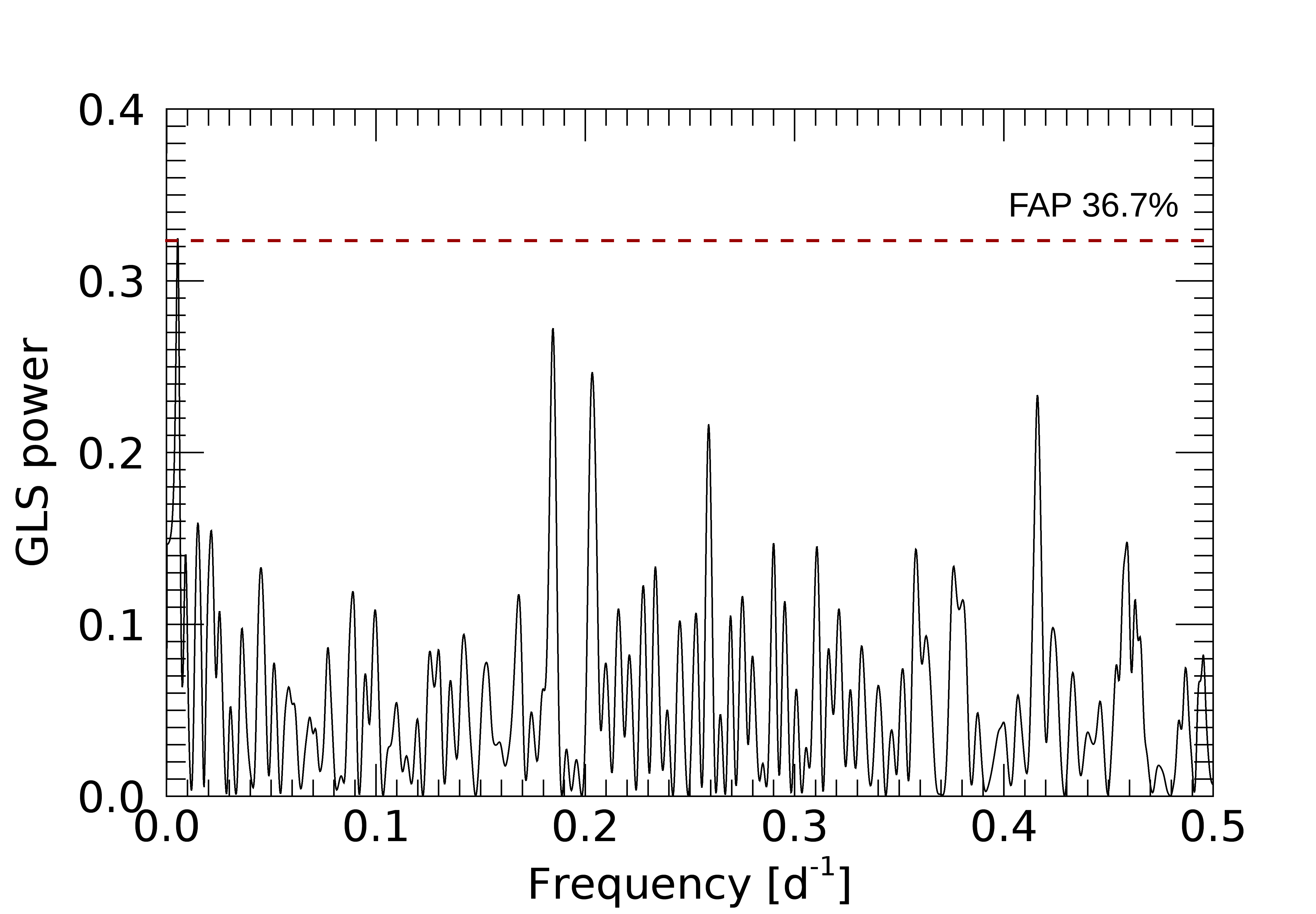}
      \caption{\textit{Upper panel}. GLS periodogram of the ESPRESSO RV residuals, after removing the best-fit Doppler signal of \object{$\pi$ Men\,b}. We assumed RV error bars with the uncorrelated jitter added in quadrature to the formal $\sigma_{\rm RV}$. The highest peak occurs at the orbital period of \object{$\pi$ Men\,c}, with a bootstrapping false alarm probability of 0.6$\%$ determined from 10\,000 simulated datasets. \textit{Middle panel}. ESPRESSO RV residuals after removing the adopted 2-planet model solution in Table\,\ref{tab:percentiles_rvmodel}. The error bars in black include the uncorrelated jitter derived from our analysis, which has been added in quadrature to the $\sigma_{\rm RV}$ uncertainties (indicated in red). \textit{Lower panel}. GLS periodogram of the ESPRESSO residuals, with the RV error bars including the uncorrelated jitter added in quadrature to the formal $\sigma_{\rm RV}$. The FAP of the main peak at $\sim$190 d was determined through a bootstrap (with replacement) analysis using 10\,000 simulated datasets. This signal is further discussed in Appendix \ref{appendixC}.}
         \label{fig:residual_period}
   \end{figure}  
  
\begin{table}
   \caption[]{Components of the proper motion vector difference for $\pi$ Men, priors, and best-fit results for the MCMC analysis of the $\Delta\mu$ time series 
   constrained by the spectroscopic orbital solution.}
          \label{tab:pimen_astro}
          \centering
         \tiny
    \begin{tabular}{l c c c}
             \hline
%             \noalign{\smallskip}
%             \textit{Derived quantities} \\
			Star name  &  Epoch & $\Delta\mu_\alpha$ & $\Delta\mu_\delta$ \\
			           &        &   (mas yr$^{-1}$)     &   (mas yr$^{-1}$) \\
			\hline
             \noalign{\smallskip}
             $\pi$ Men & Hipparcos & $0.884\pm0.398$ & $0.404\pm0.445$\\
             $\pi$ Men & Gaia & $0.591\pm0.246$ & $0.739\pm 0.263$\\
             \noalign{\smallskip}
             \noalign{\smallskip}
             \hline
%             & \multicolumn{2}{c}{Short period} & \multicolumn{2}{c}{Short period}  \\
%             Jump parameter     &  Prior & Best-fit value &  Prior & Best-fit value  \\
             Jump parameter     &  Prior & Best-fit value & \\
             \hline
             \noalign{\smallskip}
             $i_b$ [deg] & $\mathcal{U}$(0.0,180.0) & $45.8_{-1.1}^{+1.4}$ & \\
             \noalign{\smallskip}
             $\Omega_b$ [deg] & $\mathcal{U}$(0.0,360.0) & $108.8_{-0.7}^{+0.6}$ & \\
             \noalign{\smallskip}
             Mass, $m_{\rm b}$ [$\Mjup$] & (derived) & 14.1$^{+0.5}_{-0.4}$ & \\
             \noalign{\smallskip}
             \hline
      \end{tabular}
\end{table}

%__________________________________________________________________

\section{Summary and Discussion}
\label{sec:discussion}
\begin{figure}
   \centering   
  \includegraphics[width=\hsize]{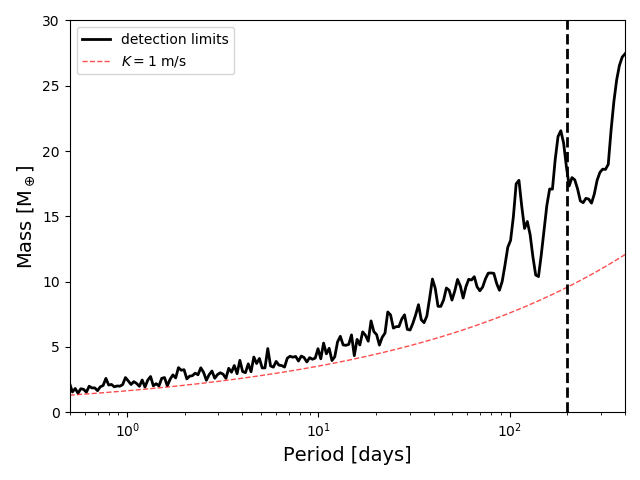}
 \caption{Detection limits for additional planetary companions in the $\pi$\,Men system. The dashed red curve corresponds to a Doppler signal with semi-amplitude of 1 \ms, and the vertical dashed line marks the location of the time span of the ESPRESSO data.}
\label{fig:detect_lim}
   \end{figure}

One main goal of our study was to assess the performance of a very high precision RV follow-up of a bright planet-hosting star with the ESPRESSO spectrograph, and the characterization of the low-mass transiting planet \object{$\pi$\,Men\,c} came as an ideal test case. The multi-planet system orbiting \object{$\pi$\,Men} was object of recent characterization studies using space-based photometry and high-precision RVs, as those collected with HARPS, thus our results can be compared with those in the literature. Figure \ref{fig:spectorbits} (lower panel) shows the low dispersion of the ESPRESSO RV measurement around the best-fit spectroscopic orbit of \object{$\pi$\,Men\,c}. After removing the best-fit Keplerian of the companion \object{$\pi$\,Men\,b} and the offsets of the \textit{pre-} and \textit{post-}technical intervention, we modelled the residuals of the ESPRESSO RVs with a Keplerian function to quantify how well the orbit of planet c is fitted using only this dataset. This Monte-Carlo analysis was performed using the open source Bayesian inference tool \textsc{MultiNest} v3.10 (e.g. \citealt{feroz13}), through the \textsc{pyMultiNest} \texttt{python} wrapper \citep{buchner14}. We obtained the Doppler semi-amplitude $K_{\rm c}$=1.5$\pm$0.3 \ms, which has basically the same precision of the value we obtained using the RVs from all the instruments. This result, based on data collected during 37 nights over a time span of 200 days, illustrates well the performance reached by ESPRESSO on such a target. From these residuals, we derived upper limits to the minimum mass of planets that may still be undetected as a function the orbital period. The detection limits are calculated by injecting trial circular orbits into the observed data (e.g. \citealt{cumming99}). We explore orbital periods from 0.5 days to twice the timespan of the ESPRESSO data, and semi-amplitudes up to 10 \kms (using a binary search). For ten linearly spaced phases, the periodogram power at the injected period is compared with the 1$\%$ false alarm probability (FAP) level in the original residuals. If, for all phases, the former is higher, the injected planet is considered detected. Using our measured stellar mass, the semi-amplitudes are converted to the minimum mass of the planet. The detection limits are shown in Fig.\,\ref{fig:detect_lim}, from which we can conclude that undetected companions to \object{$\pi$ Men} within the orbit of \object{$\pi$ Men\,c} should have a minimum mass lower than $\sim$2 $M_\oplus$. We can exclude planets with minimum mass less than $\sim$10 $\mearth$ and with orbital period greater than 100 days.

\begin{figure}
   \centering
   \includegraphics[width=\hsize]{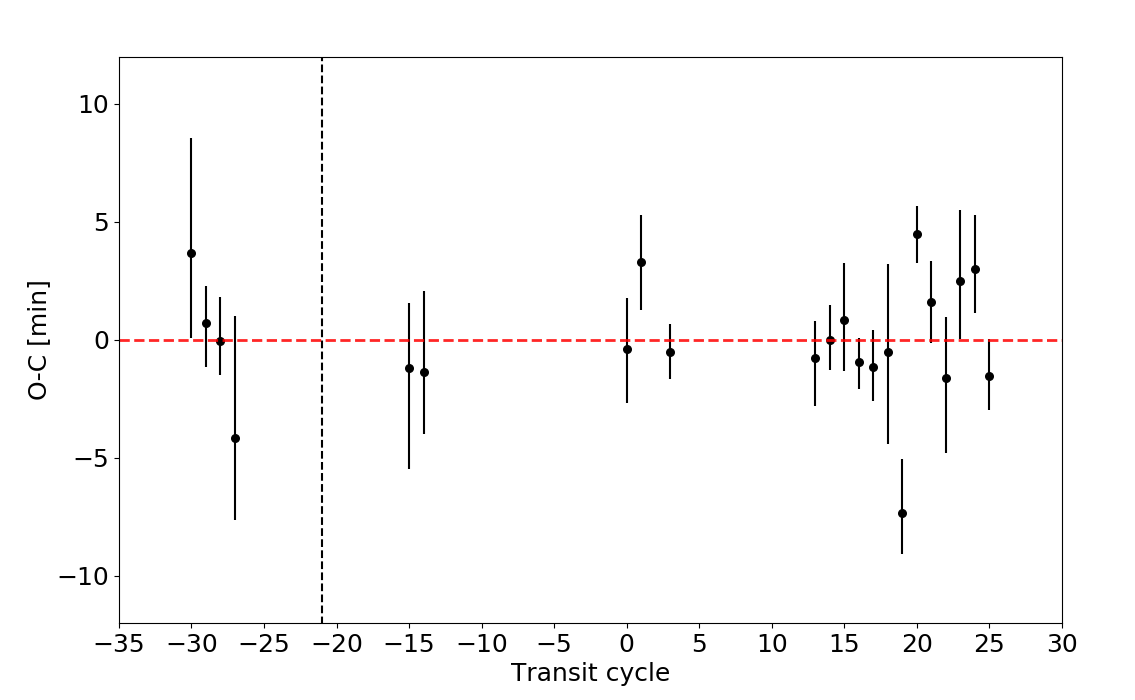}
      \caption{Transit timing variations measured from the TESS light curve. The dashed vertical line shows the epoch of the periastron passage of $\pi$\,Men\,b.
              }
         \label{fig:ttv}
   \end{figure}

Our analysis of HD39091 based on spectroscopy and photometry was enhanced by the use of astrometry, that allowed for a much more complete characterization of the system's properties and architecture. The evidence for a large mutual inclination between \object{$\pi$\,Men}\,b and c naturally raises the possibility for strong secular dynamical perturbations on the inner planet's orbital arrangement. In the co-planar case, the amplitude of TTVs for \object{$\pi$\,Men\,c} is not expected to exceed a few tens of seconds (e.g., \citealt{holman05}). No clear departure from a linear ephemeris is detected within the error bars in the residuals of the time of transit centre (Fig.\,\ref{fig:ttv}). We used \textsc{TTVFast}\citep{Deck2014} to explore whether the high mutual inclination might induce a detectable signal, also given the fact that the periastron passage of \object{$\pi$\,Men\,b} occurred within the time span of the \textit{TESS} observations. However, the analysis with \textsc{TTVFast}, limited to the time span of the \textit{TESS} observations, did not produce any detectable TTV signal and is in agreement with our observations, resulting in the longitude of the ascending node of \object{$\pi$\,Men\,c} fully unconstrained. This is not surprising given the fact that only a small fraction of the full orbit of the perturber has been covered, a circumstance that could prevent recovering a TTV signal \citep{Deck2014}. 

The high value of $i_\mathrm{rel}$ could cause in principle significant secular evolution in eccentricity and inclination of \object{$\pi$\,Men\,c} (see e.g. \citealt{kozai1962,lidov1962,holman1997}), which might then shift out of the transiting configuration observed today. The system's architecture is also suggestive of a violent dynamical evolution history, that might point to a high-eccentricity migration scenario and a significant degree of spin-orbit misalignment of the transiting inner planet (see e.g. \citealt{fabrycky2007,chatte2008,ogilvie2014,hamers2017}).  

Based on the deuterium burning-mass limit for separating planets and brown dwarfs, which is theoretically established at $\sim$13 $\Mjup$ for solar metallicity and the cosmic abundance of deuterium \citep{burrows95,saumon96,chabrier00}, \object{$\pi$\,Men\,b} should be classified as a brown dwarf. Brown dwarf companions to a main sequence star appear to be very rare in close orbits ($<$3 AU), and their occurrence rate is much lower than for giant planets and stars. For instance, \cite{grether06} found that $\sim$16$\%$ of Sun-like stars have close companions (orbital period $<$5 yr) more massive than Jupiter, but only $<1\%$ are brown dwarfs, while 11$\%$ and 5$\%$ are stars and giant planets, respectively. This paucity is traditionally refereed to as the \textit{brown dwarf desert}. Today, the $Kepler/K2$ and $TESS$ missions and the $superWASP$ survey have proved that this desert is not so ``dry'' as originally thought. Several brown dwarfs have been detected with short orbital periods (see, e.g., \citealt{carmichael19,persson19,charmichael20,subjak20,parviainen20}), and their occurrence rate has been revised to 2.0$\pm$0.5$\%$ by \cite{kiefer19}. Our precise mass determination for $\pi$ Men\,b contributes to further populate the brown dwarf desert.

The results of our study, based on multi-technique observations, make $\pi$ Men a benchmark multi-body system -- with a brown dwarf cohabiting with a super-Earth around a solar-like star -- for which the 3-D architecture has been unveiled with precision. This indeed encourages further follow-up and detailed modeling of the \object{$\pi$\,Men} planetary system to understand its formation and evolution. We note that the nominal schedule of future \textit{TESS} observations available at the moment includes the field of \object{$\pi$\,Men} for six more sectors in 2020-2021. The new data could be used to further constrain the TTVs of planet c.

\textit{Note}. Few days before the formal acceptance of this paper, an independent study about the architecture of the \object{$\pi$\,Men} planetary system was published \citep{xuan20}. The results of that work, based on public data and not including the ESPRESSO observations, confirm the high mutual inclination of the orbital planes of \object{$\pi$\,Men\,b and c}. Our results are in agreement with those of \cite{xuan20} and are characterized by a better formal precision.

\begin{acknowledgements}
We thank the referee Pierre Kervella for his comments that improved the contents of this work. 
M.D. acknowledges financial support from Progetto Premiale 2015 FRONTIERA (OB.FU. 1.05.06.11) funding scheme of the Italian Ministry of Education, University, and Research, and also acknowledge the computing centres of INAF - Osservatorio Astronomico di Trieste/Osservatorio Astrofisico di Catania, under the coordination of the CHIPP project, for the availability of computing resources and support. C.L. and F.P. would like to acknowledge the Swiss National Science Foundation (SNSF) for supporting research with ESPRESSO through the SNSF grants nr. 140649, 152721, 166227 and 184618. The ESPRESSO Instrument Project was partially funded through SNSF’s FLARE Programme for large infrastructures. This work has been carried out in part within the framework of the NCCR PlanetS supported by the Swiss National Science Foundation.
This work was supported by FCT - Fundação para a Ciência e a Tecnologia through national funds and by FEDER through COMPETE2020 - Programa Operacional Competitividade e Internacionalização by these grants: UID/FIS/04434/2019; UIDB/04434/2020; UIDP/04434/2020; PTDC/FIS-AST/32113/2017 \& POCI-01-0145-FEDER-032113; PTDC/FIS-AST/28953/2017 \& POCI-01-0145-FEDER-028953; PTDC/FIS-AST/28987/2017 \& POCI-01-0145-FEDER-028987. S.C.C.B., V.Z.A., S.G.S. and J.P.S.F. acknowledge support from FCT through FCT contracts nr. IF/01312/2014/CP1215/CT0004, IF/00650/2015/CP1273/CT0001, IF/00028/2014/CP1215/CT0002, DL 57/2016/CP1364/CT0005.
O.D.S.D. is supported in the form of work contract (DL 57/2016/CP1364/CT0004) funded by national funds through Funda\,c\~ao para a Ci\^encia e Tecnologia (FCT).
J.L.-B. has been funded by the Spanish State Research Agency (AEI) Projects No.ESP2017-87676-C5-1-R and No. MDM-2017-0737 Unidad de Excelencia "Mar\'ia de Maeztu"- Centro de Astrobiolog\'ia (INTA-CSIC).
J.P.F. is supported in the form of a work contract funded by national funds through FCT with reference DL57/2016/CP1364/CT0005.
V.B. acknowledges support by the Swiss National Science Foundation (SNSF) in the frame of the National Centre for Competence in Research ``PlanetS''.
This project has received funding from the European Research Council (ERC) under the European Union’s Horizon 2020 research and innovation programme (project Four Aces, grant agreement No 724427).
R.R., C.A.P., J.I.G.H., A.S.M., and M.R.Z.O. acknowledge financial support from the Spanish Ministry of Science and Innovation (MICINN) projects AYA2017-86389-P and AYA2016-79425-C3-2-P.
J.I.G.H. acknowledges financial support from the Spanish MICINN under the 2013 Ram\'on y Cajal program RYC-2013-14875.
M.T.M. thanks the Australian Research Council for \textsl{Future Fellowship} grant FT180100194 which supported this work.
Partial financial support from the agreement ASI-INAF n.2018-16-HH.0 is gratefully acknowledgded.
This work has made use of data from the European Space Agency (ESA) mission {\it Gaia} (\url{https://www.cosmos.esa.int/gaia}), processed by the {\it Gaia} Data Processing and Analysis Consortium (DPAC, \url{https://www.cosmos.esa.int/web/gaia/dpac/consortium}). Funding for the DPAC has been provided by national institutions, in particular the institutions participating in the {\it Gaia} Multilateral Agreement.
This publication makes use of The Data $\&$ Analysis Center for Exoplanets (DACE), which is a facility based at the University of Geneva (CH) dedicated to extrasolar planets data visualisation, exchange and analysis. DACE is a platform of the Swiss National Centre of Competence in Research (NCCR) PlanetS, federating the expertise in Exoplanet research. The DACE platform is available at \url{https://dace.unige.ch}.

\end{acknowledgements}

\bibliographystyle{aa} % style aa.bst
\bibliography{references.bib} % your references Yourfile.bib

\begin{appendix}

\section{Coadded ESPRESSO spectrum}

\begin{figure*}
   \centering
   \includegraphics[width=0.6\hsize, angle=90]{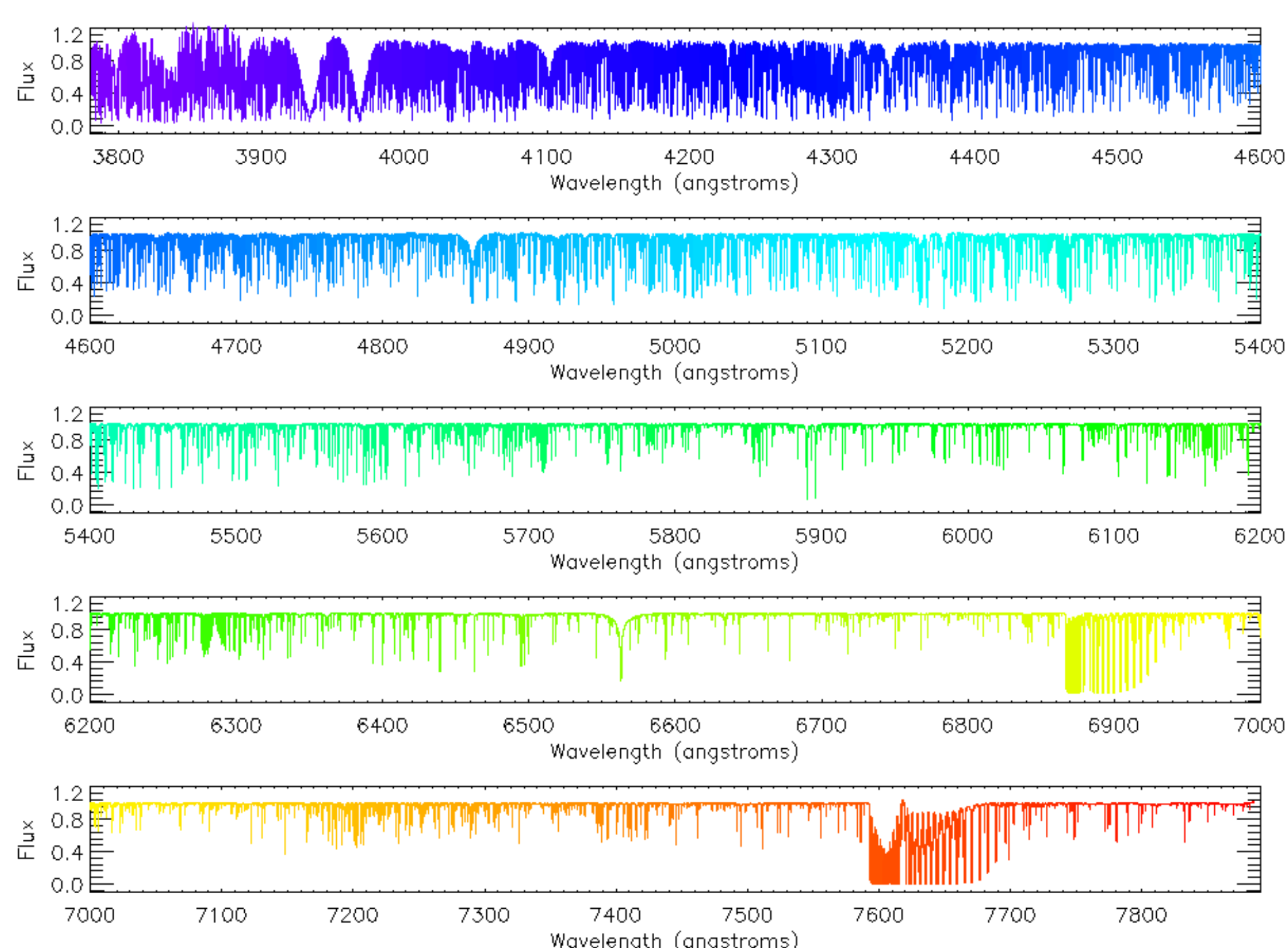}\\
   \vspace*{+10mm}
   \includegraphics[width=0.85\hsize]{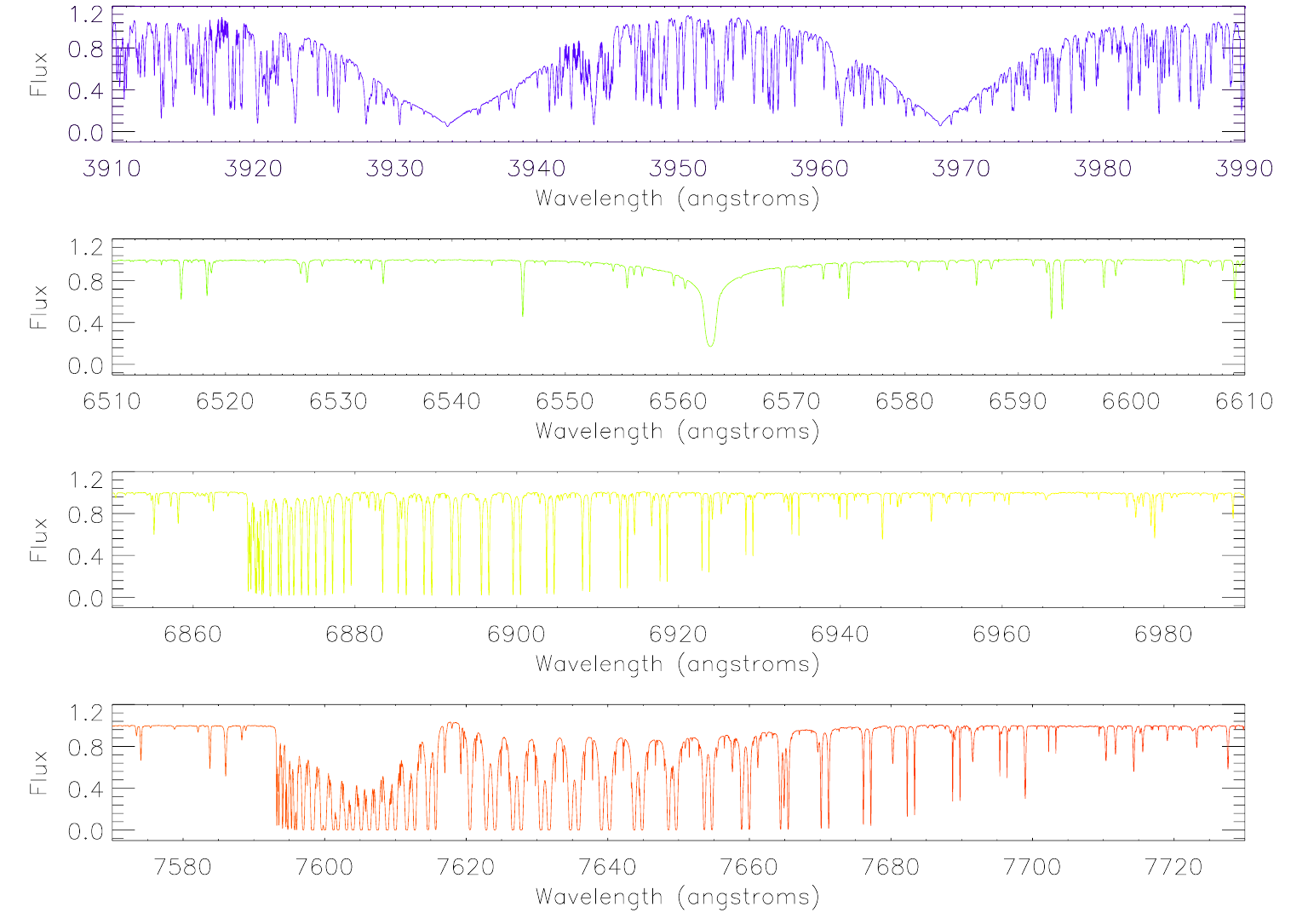}
      \caption{\textit{Upper plot.} Coadded, normalized, merged, RV-corrected ESPRESSO spectrum of \object{$\pi$ Men} using the starII DAS workflow in a subset of 40 ESPRESSO spectra. \textit{Lower plot.} Selected regions of the spectrum, to illustrate its quality.
              }
         \label{fig:coaddedspectrum}
   \end{figure*}

\clearpage

\section{ESPRESSO and CORALIE data}

The radial velocities and activity diagnostics used in this work are listed in Table \ref{data_pimensae} and \ref{data_pimensae_2}.
The data are made publicly available also from the DACE platform through the Web page  \url{https://dace.unige.ch/radialVelocities/?pattern=Pi%20Men} or the \texttt{python} API's \url{https://dace.unige.ch/radialVelocities/?pattern=Pi%20Men}. 

\onecolumn
%\onllongtab{
%\begin{landscape}
\begin{table}
\caption{Radial velocities and activity indicators of $\pi$\,Men extracted from ESPRESSO spectra with the version 2.0.0 of the DRS pipeline.} 
\label{data_pimensae}
\begin{tabular}{ccccccccc}
\hline
\hline
Time  & RV & $\sigma_{\rm RV}$ & FWHM & BIS & S-index & S-index error & H$_\alpha$-index & H$_\alpha$-index error \\
(BJD-2\,450\,000) & (\ms)  & (\ms)  & (\ms)  & (\ms)  & & & \\
\hline
8367.869454 & 10874.79 & 0.26 & 8726.89 & -0.0350& 0.157607 & 0.000029 & 0.190834 & 0.000014 \\
8367.871444 & 10876.16 & 0.24 & 8727.41 & -0.0349& 0.158518 & 0.000025 & 0.191048 & 0.000012 \\
8367.873341 & 10875.78 & 0.25 & 8726.28 & -0.0347& 0.157664 & 0.000026 & 0.190987 & 0.000013 \\
8367.875278 & 10875.10 & 0.22 & 8728.95 & -0.0347& 0.158364 & 0.000021 & 0.190946 & 0.000011 \\
... 		& ... 	   & ...  & ...     &  ...   & ...      & ...      & ...      & ...   \\
\hline
\hline
\end{tabular}
\end{table}

%\twocolumn

\begin{table}
\caption[]{Radial velocities of $\pi$\,Men extracted from CORALIE spectra.} 
\label{data_pimensae_2}
\centering
\begin{tabular}{ccc}
\hline
\hline
Time  & RV & $\sigma_{\rm RV}$ \\
(BJD-2\,450\,000) & (\ms) & (\ms) \\
\hline
CORALIE-98 \\
\hline
1131.806114 & 10609.63 &  8.74 \\
1139.791656 & 10592.62 &  6.14 \\
1189.666147 & 10617.22 &  5.81 \\
1256.511024 & 10625.82 &  5.32 \\
1453.866446 & 10584.39 &  5.70 \\
... & ... & ...\\
\hline\hline
\end{tabular}
\end{table}

% \twocolumn

\section{Further analysis of the ESPRESSO RV residuals}
\label{appendixC}

\begin{figure}
   \centering
   \includegraphics[width=\hsize]{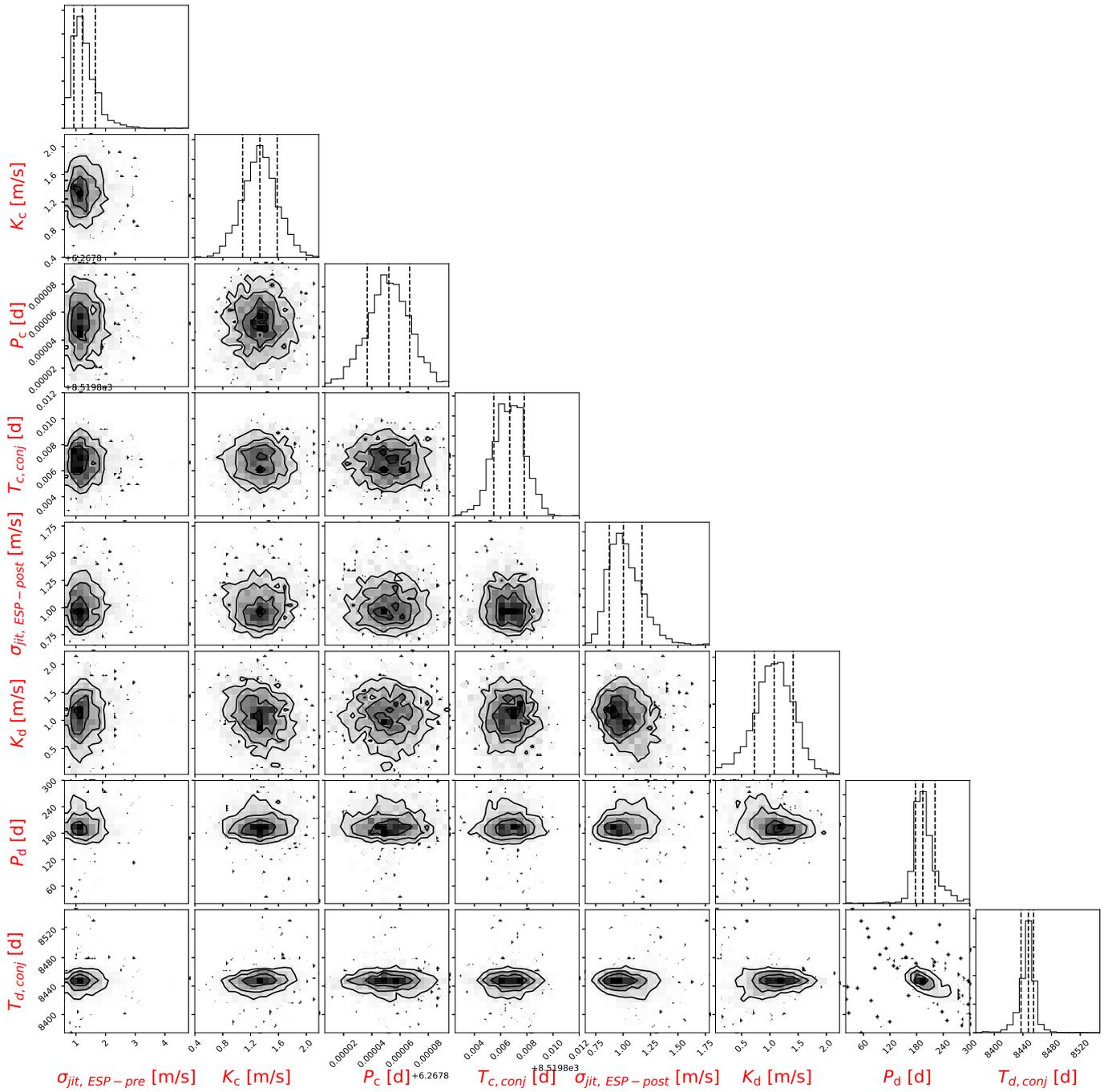} \\
      \caption{Posterior distributions for the free parameters of the 2-planet model tested on the RV ESPRESSO residuals, after removing the spectroscopic orbit of $\pi$ Men\,b and instrumental offsets from the orginal dataset.}
         \label{fig:posterior_d}
\end{figure}

\begin{figure}
   \centering
   \includegraphics[width=\hsize]{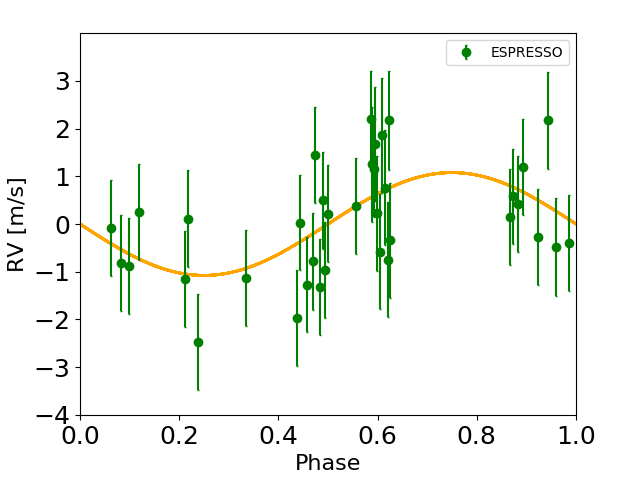} \\
      \caption{Best-fit model (orange curve) for the additional $\sim$190-d signal found in the RV residuals of ESPRESSO, as derived with a Monte-Carlo analysis. The error bars include a constant jitter term added in quadrature to the formal RV uncertainties.}
         \label{fig:signal_d}
\end{figure}

As discussed in Sect. \ref{sec:comb_analysis}, the GLS periodogram of the 2-planet model RV residuals shows a peak at $\sim$190 d (last panel of Fig. \ref{fig:residual_period}). Even though it appears not significant according to a bootstrap analysis, nonetheless we investigated the properties of this signal in more detail by performing a Monte-Carlo analysis with \textsc{MultiNest} and a Bayesian model comparison. 
To this purpose, we considered the ESPRESSO RV residuals after subtracting the best-fit spectroscopic orbit of \object{$\pi$\,Men\,b} and instrumental offsets, and we modelled them with two Keplerians setting their eccentricities to zero, one for the Doppler signal due to \object{$\pi$\,Men\,c}, and using uninformative priors for the semi-amplitude, period and time of inferior conjunction of the other Doppler signal ($K: \mathcal{U}(0,3) \ms$; $P: \mathcal{U}(0,300)$ d; $T_{0,c}: \mathcal{U}(2\,458\,370,2\,458\,690)$ BJD). We found that the fit improves with respect to including only $\pi$ Men\,c (see Sect. \ref{sec:discussion}). The Bayesian factor is $\sim+5$, favouring the model with 2 signals, with semi-amplitude $K_c$=1.3$\pm$0.2 $\ms$ (slightly lower and more significant than $K_c$=1.5$\pm$0.3 \ms), and the uncorrelated jitter of the post-intervention data slightly decreases to 1.0$^{+0.2}_{-0.1}$ $\ms$. For the second signal, we found $K_{d}=1.1^{+0.3}_{-0.4} \ms$ and $P_{d}=194^{+28}_{-17}$ d. The posterior distributions for the free model parameters are shown in Fig. \ref{fig:posterior_d}, and the best-fit model for the additional signal is shown in Fig. \ref{fig:signal_d}. If due to a third planet in the system, this signal would correspond to a minimum mass of $\sim$10 \mearth. The results of our analysis are not sufficiently robust to reach any clear conclusion about the nature of this signal that appears in the ESPRESSO data, and we did not perform any dynamical analysis to verify the orbital stability. Further spectroscopic follow-up is indeed required to confirm the signal and improve the phase coverage.

\section{Radial velocity and astrometric joint analysis}
\label{appendixD}

\begin{figure}
   \centering
   \includegraphics[width=\hsize]{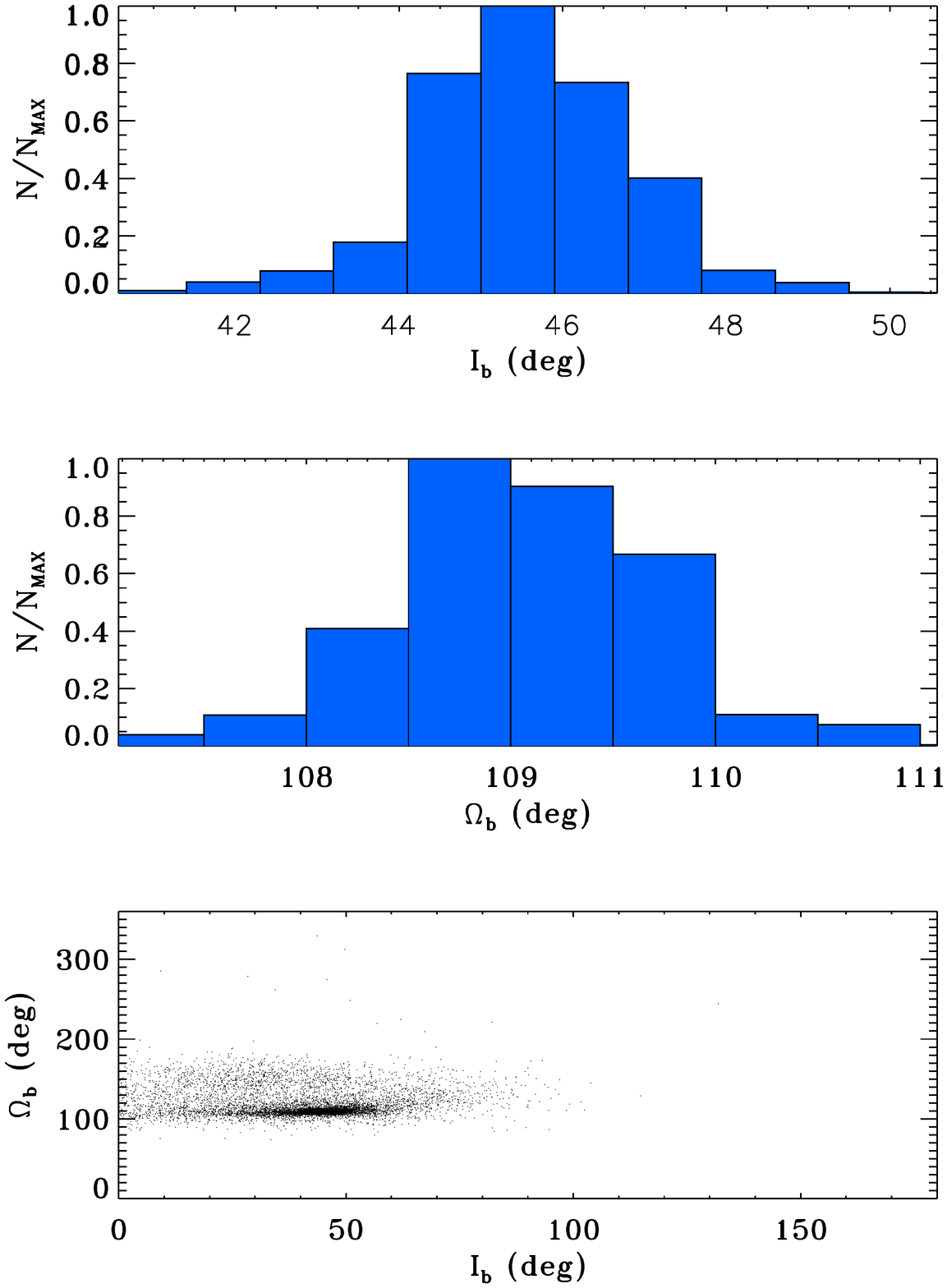} \\
      \caption{\textit{Top and central panels}: posterior distributions for $i_b$ and $\Omega_b$. \textit{Lower panel}: joint posterior distributions for the two model parameters.}
         \label{fig:posteriors_astro}
\end{figure}

\begin{figure}
   \centering
   \includegraphics[width=\hsize]{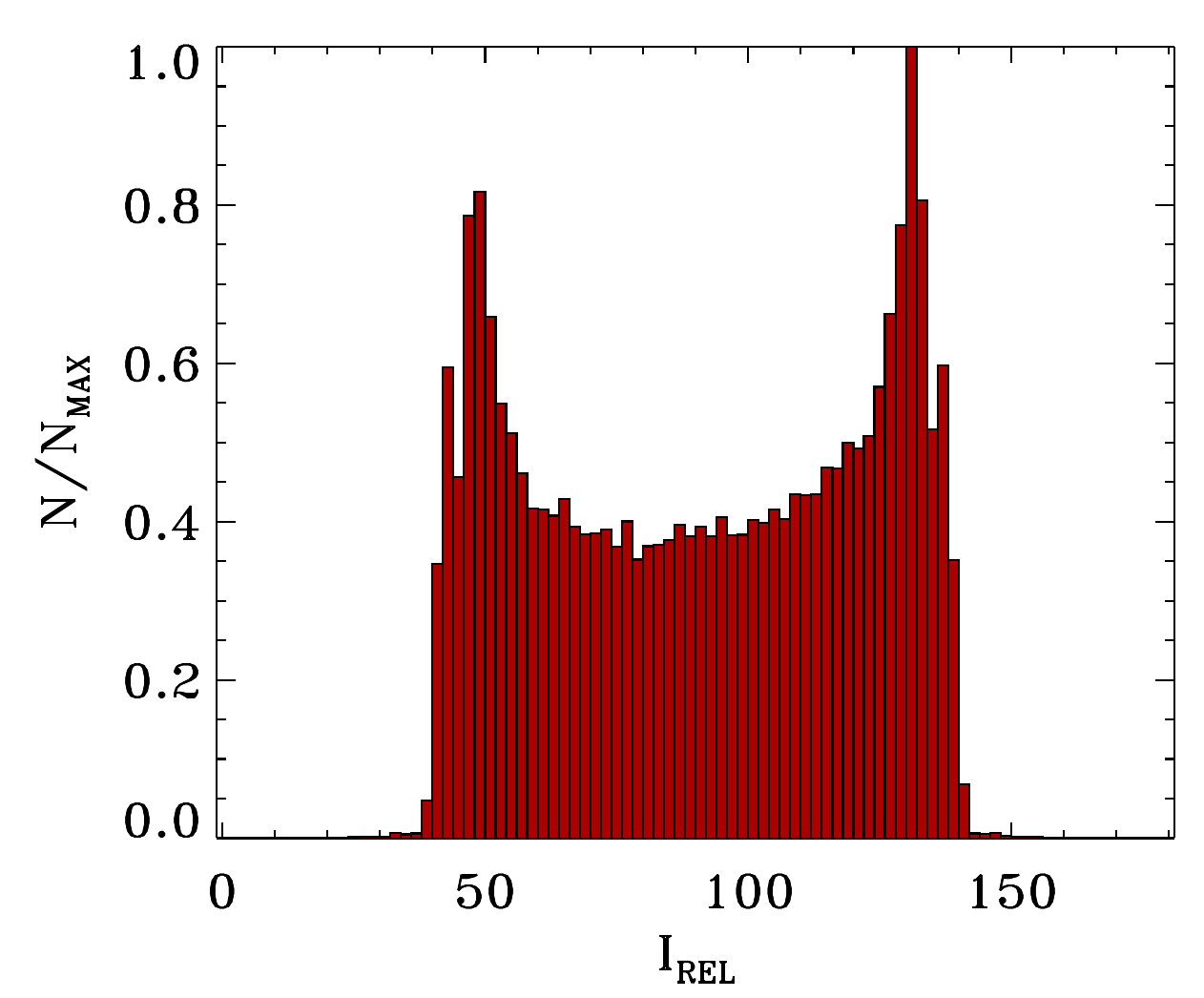} \\
      \caption{Distribution of possible mutual inclinations between $\pi$ Men b and c.}
         \label{fig:rel_inc}
\end{figure}

Fig. \ref{fig:posteriors_astro} shows the posterior distributions for $i_b$ and $\Omega_b$ derived from the analysis described in Sect. \ref{sec:astrometry}. As one can see, the orbital plane inclination angle $i_{\rm b}$  and the longitude of the ascending node $\Omega_{\rm b}$ for planet b are both fitted with high formal precision: $i_b=45.8_{-1.1}^{+1.4}$\,deg and $\Omega_b=108.8_{-0.7}^{+0.6}$\,deg. 
Fig. \ref{fig:rel_inc} shows the distribution of the mutual inclination angles $i_\mathrm{rel}$ between the companions b and c to \object{$\pi$ Men}.

%\section{Details of the Hipparcos/Gaia DR2 astrometric analysis of $\pi$ Men }
%\label{appendixC}

\end{appendix}

\end{document}